\documentclass{aa}
\usepackage{graphicx}

\begin{document}

\title{Precession, nutation, and space geodetic determination of the Earth's variable gravity field}

\author{G. Bourda \and N. Capitaine}

\offprints{G. Bourda, \email{Geraldine.Bourda@obspm.fr}}

\institute{Observatoire de Paris, SYRTE/UMR 8630-CNRS, \\ 
61 avenue de l'Observatoire, 75014 Paris, France, 
email: bourda@syrte.obspm.fr}

\date{Received: 25 June 2004, Accepted: 10 August 2004}

\abstract{Precession and nutation of the Earth depend on the Earth's dynamical flattening, $H$, which is closely related to the second degree zonal coefficient, $J_2$ of the geopotential. A small secular decrease as well as seasonal variations of this coefficient have been detected by precise measurements of artificial satellites (Nerem et al. 1993, Cazenave et al. 1995) which have to be taken into account for modelling precession and nutation at a microarcsecond accuracy in order to be in agreement with the accuracy of current VLBI determinations of the Earth orientation parameters. However, the large uncertainties in the theoretical models for these $J_2$ variations (for example a recent change in the observed secular trend) is one of the most important causes of why the accuracy of the precession-nutation models is limited (Williams 1994, Capitaine et al. 2003). We have investigated in this paper how the use of the variations of $J_2$ observed by space geodetic techniques can influence the theoretical expressions for precession and nutation. We have used time series of $J_2$ obtained by the ``Groupe de Recherches en G\'eod\'esie spatiale'' (GRGS) from the precise orbit determination of several artificial satellites from 1985 to 2002 to evaluate the effect of the corresponding constant, secular and periodic parts of $H$ and we have discussed the best way of taking the observed variations into account. We have concluded that, although a realistic estimation of the $J_2$ rate must rely not only on space geodetic observations over a limited period but also on other kinds of observations, the monitoring of periodic variations in $J_2$ could be used for predicting the effects on the periodic part of the precession-nutation motion.

}

\titlerunning{Precession, nutation, and Earth variable gravity field}

\maketitle


\section{Introduction}

Expressions for the precession of the equator rely on values for the precession rate in longitude that have been derived from astronomical observations (i.e. observations that were based upon optical astrometry until the IAU~1976 precession, and then on Very Long Baseline Interferometry (VLBI) observations for more recent models). The IAU~2000 precession-nutation model provided by Mathews et~al. (2002) (denoted MHB~2000 in the following), that was adopted by the IAU beginning on 1~January~2003, includes a new nutation series for a non-rigid Earth and corrections to the precession rates in longitude and obliquity that were estimated from VLBI observations during a 20-year period. The precession in longitude for the equator being a function of the Earth's dynamical flattening $H$, observed values for this precession quantity are classically used to derive a realistic value for $H$. Such a global dynamical parameter of the Earth is generally considered as a constant, except in a few recent models for precession (Williams 1994, Capitaine et al. 2003) or nutation (Souchay \& Folgueira 1999, Mathews et~al. 2002, Lambert \& Capitaine 2004) in which either the secular or the zonal variations of this coefficient are explicitly considered through simplified models.

The recent implementation of the IAU~2000A precession-nutation model guarantees an accuracy of about 200 $\mu$as in the nutation angles, and all the predictable effects that have amplitudes of the order of 10~$\mu$as have therefore to be considered. One of these effects is the influence of the variations ($\Delta H$) in the Earth's dynamical flattening, which are not explicitly considered in the IAU~2000A precession-nutation model. Furthermore, the IAU~2000 precession is based on an improvement of the precession rates values derived from recent VLBI measurements, but it does not improve the higher degree terms in the polynomials for the precession angles $\psi_A$, $\omega_A$ of the equator (see Fig. \ref{fig:Prec_Nut_Angles}). This precession model is not dynamically consistent because the higher degree precession terms are actually dependent on the precession rates (Capitaine et~al.,~2003) and need to be improved, even though VLBI observations are unable to discriminate between recent solutions due to the limited span of the available data (Capitaine et al., 2004). One alternative way for such an improvement is to improve the model for the geophysical contributions to the precession angles and especially the influence of $\Delta H$ (or equivalently $\Delta J_2$). 

The $H$ parameter is linked to the dynamical form-factor, $J_2$ for the Earth (i.e.~the $C_{20}$ harmonic coefficient of the geopotential) which is determined by space geodetic techniques on a regular basis. Owing to the accuracy now reached by these techniques, the temporal variation of a few Earth gravity field coefficients, especially $\Delta C_{20}$, can be determined (for early studies, see for example Nerem et al.~(1993), Cazenave et al.~(1995) or Bianco et al.~(1998)). They are due to Earth oceanic and solid tides, as well as mass displacements of geophysical reservoirs and post-glacial rebound for $\Delta C_{20}$. This coefficient $C_{20}$ can be related to the Earth's orientation parameters and more particularly to the Earth precession-nutation, through $H$. The purpose of this paper is to use space geodetic determination of the geopotential to estimate $\Delta H$, in order to investigate its influence on the precession-nutation model. The $C_{20}$ data used in this study have been obtained from the positioning of several satellites between 1985 and 2002. We estimate also the constant part of $H$, based on such space geodetic measurements, and compare its value and influence on precession results with respect to those based on VLBI determinations. 

In Sect.~2 we recall the equations expressing the equatorial precession angles as a function of the dynamical flattening $H$. We provide the numerical values implemented in our model, compare the values obtained for $H$ by various studies and discuss the methods on which they rely. In Sect.~3 the relationship between $\Delta H$ and $\Delta C_{20}$ is discussed, depending on the method implemented. We explain how these geodetic data are taken into account in Sect.~4. We present our results in Sect.~5, and discuss them in the last part. We investigate how the use of a geodetic determination of the variable geopotential can influence the precession-nutation results, considering first the precession alone, and second the periodic contribution. 

In the whole study, the time scale for $t$ is TT Julian centuries since J2000, which will be denoted cy.


\section{Theoretical effect of $\Delta H$ on precession}

This section investigates the theoretical effect of the variations $\Delta H$ in the Earth's dynamical flattening on the precession expressions.

\subsection{Relationship between $H$ and the precession of the equator}

\begin{figure}
\centering
\includegraphics[width=8cm]{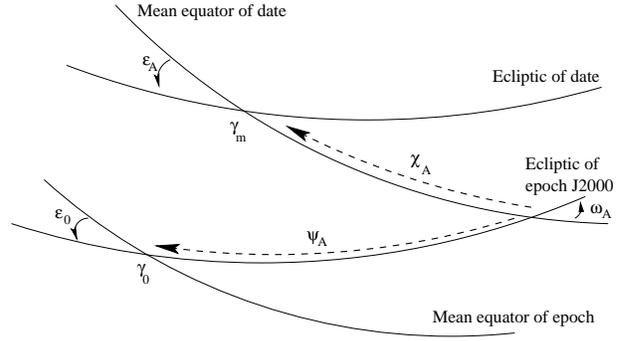}
\caption{Angles $\psi_A$ and $\omega_A$ for the precession of the equator: $\gamma_m$ is the mean equinox of the date and $\gamma_0$ is the equinox of the epoch J2000.0.}
\label{fig:Prec_Nut_Angles}
\end{figure}

The two basic angles $\psi_A$ and $\omega_A$ (see Fig.~\ref{fig:Prec_Nut_Angles}) for the precession of the equator are provided by the following differential equations (see Eq.~(29) of Williams (1994) or Eq.~(24) of Capitaine et~al. (2003)):
\begin{eqnarray}\label{eq:Syst_Prec_Nut}
\sin \omega_A ~\frac{d\psi_A}{dt} & = & \left( r_{\psi} ~\sin \epsilon_A \right) ~\cos \chi_A - r_{\epsilon} ~\sin \chi_A   \nonumber  \\
\frac{d\omega_A}{dt} & = & r_{\epsilon} ~\cos \chi_A + \left( r_{\psi} ~\sin \epsilon_A \right) ~\sin \chi_ A
\end{eqnarray}
where $r_{\psi}$ and $r_{\epsilon}$ are respectively the precession rates in longitude and obliquity, $\epsilon_A$ is the obliquity of the ecliptic of date and $\chi_A$ the planetary precession angle, determining the precession of the ecliptic. Updated expressions for these precession quantities are given in Capitaine et~al.~(2003). An expression for the precession rates, $r_{\psi}$ in longitude and $r_{\epsilon}$ in obliquity, is provided in detail in Williams (1994) and Capitaine et al.~(2003) as a function of various contributions. The precession rate in longitude can be written as $r_{\psi} = r_0 + r_1 ~t + r_2 ~t^2 + r_3 ~t^3$ where the largest first-order term in $r_0$ is the luni-solar contribution denoted ${f_{01}}_{|_{LS}} \cos\epsilon_0$, where $\epsilon_0$ is the obliquity of the ecliptic at J2000. It is such that (Kinoshita 1977, Dehant \& Capitaine 1997): 
\begin{equation}\label{eq:luni_solar_effect}  
{f_{01}}_{|_{LS}} = k_m ~M_0 + k_s ~S_0
\end{equation}
in which $M_0$ and $S_0$ are the amplitudes of the zero-frequency Moon and Sun attractions, respectively, and:
\begin{eqnarray}  
k_m & = & 3 ~H ~\frac{m_m}{m_{m} + m_{\oplus}} ~\frac{1}{{F_2}^3} ~\frac{n_{m}^2}{\Omega} = H ~K_m   \label{eq:Km}\\
k_s & = & 3 ~H ~\frac{m_{\odot}}{m_{\odot}+m_{m} + m_{\oplus}} ~\frac{n_{\odot}^2}{\Omega} = H ~K_s  \label{eq:Ks} 
\end{eqnarray}
In the above expressions, $H$ is the Earth's dynamical flattening, $m_m$, $m_{\odot}$ and $m_{\oplus}$ are the masses of the Moon, the Earth and the Sun, respectively, $n_m$ is the Moon mean motion around the Earth, $n_{\odot}$ the Earth mean motion around the Sun, $\Omega$ is the mean angular velocity of the Earth and $F_2$ a factor for the mean distance of the Moon. Current numerical values for such a problem are (Souchay \& Kinoshita, 1996):
\begin{eqnarray}  
& & M_0  =  496303.66 \times 10^{-6}   \nonumber \\
& & S_0  =  500210.62 \times 10^{-6}   \nonumber \\
& & k_m  =  7546''.7173289~\mbox{/cy}     \\
& & k_s  =  3475''.1883295~\mbox{/cy}    \nonumber \\
& & {f_{01}}_{|_{LS}} \cos\epsilon_0  =  5040''.6445 ~\mbox{/cy}    \nonumber 
\end{eqnarray}
and (see Kinoshita, 1977):
\begin{eqnarray}  
& & F_2 = 0.999093142   \nonumber  
\end{eqnarray}

Hence, the link between the precession of the equator ($\psi_A$ and $\omega_A$ angles) and the Earth's dynamical flattening ($H$) is shown by Eq.~(\ref{eq:Syst_Prec_Nut}), Eq.~(\ref{eq:luni_solar_effect}), Eq.~(\ref{eq:Km}), (\ref{eq:Ks}) and Eq.~(\ref{eq:r0}) of Sect. 2.2. Classically, $H$ is related to ${f_{01}}_{|_{LS}}$ derived from observations by:
\begin{equation}\label{eq:H}
H = \frac{{f_{01}}_{|_{LS}} }{K_m~M_0 + K_s~S_0}
\end{equation}


\subsection{Astronomical determination of $H$}

We can write $r_0$ as:
\begin{eqnarray}\label{eq:r0}
r_0 & = & {f_{01}}_{|_{LS}} \cos \epsilon_0 + {f_{01}}_{|_{PL}} \cos \epsilon_0            \\
    & + & H \times \mbox{lunisolar second order effects}                         \nonumber \\
    & + & H \times \left( J_2 ~\mbox{and planetary} \right) ~\mbox{tilt effects} \nonumber \\
    & + & J_4 ~\mbox{lunisolar effect}                                           \nonumber \\
    & - & \mbox{geodesic precession}                                             \nonumber \\ 
    & + & \mbox{non-linear effects (Mathews et al., 2002)}                       \nonumber 
\end{eqnarray}
where ${f_{01}}_{|_{PL}}$ is the first order term of the planetary contribution (also proportional to $H$).
Classically, H is derived from observationally determined values of $r_0$. The measurement of $r_0$ should be corrected by removing the modelled contributions other than the lunisolar first order effect (see Eq.~(\ref{eq:r0})). Hence, we obtain a value for ${f_{01}}_{|_{LS}}$, which is the only term with sufficiently large amplitude (of the order of $5000 ''$/cy) to be sensitive to small changes in the value of the dynamical ellipticity $H$ of the Earth (see Eq.~(\ref{eq:H})). So, given the other contributions provided by the theory, we can derive the value of $H$ from the observed value of $r_0$ and the model for the lunisolar first order effects.

A major problem consists in choosing the constant value of $H$. Indeed, depending on the authors, it differs by about $10^{-7}$ (Table \ref{tab:Hd_constant}). This is due to the different measurements and models implemented (see Fig.~1 of Dehant \& Capitaine (1997); Fig.~5 of Dehant et~al.~(1999)).
On the one hand the optical measurements give values of the general precession in longitude $p_A$ referred to the ecliptic of date, whereas VLBI gives measurements relative to space.
On the other hand, the various constants and models used for obtaining the value for $H$ from a measured value (optical, Lunar laser ranging or VLBI) are different depending on the study considered (see Eq.~(\ref{eq:r0})). 

Classically, $\psi_A$ is developed in a polynomial form of $t$ as: $\psi_A = \psi_0+\psi_1~t+\psi_2~t^2 +\psi_3~t^3$. In Table \ref{tab:Hd_constant}, we recall the different values used (i) for $\psi_1$ (i.e. the precession rate in longitude, $\psi_1 = r_0$), directly obtained from VLBI measurements, and (ii) for $p_1$ which is the observationally determined value of precession in the optical case: $\psi_1 = p_1 + \chi_1 ~cos~\epsilon_0$ (Lieske et al., 1977).

\begin{table*}[htbp]
\centering
\caption{Comparison between constants used for different determinations of the dynamical flattening ($H$): (1) the precession rate in longitude ($\psi_1$), (2) the speed of the general precession in longitude ($p_1$), (3) the geodesic precession ($p_g$) and (4) the obliquity of the ecliptic at J2000.0 ($\epsilon_0$). The observational value actually used for each study is written in bold.}\label{tab:Hd_constant}
\begin{tabular}{llllll}
\hline
\hline
\noalign{\smallskip}
                            &                  &  (1)              &   (2)           &  (3)      &  (4)                    \\
Sources                     &     $H$          &  $\psi_1$         &   $p_1$         &  $p_g$    &  $\epsilon_0$           \\
                            & ($\times ~10^3$) & \multicolumn{3}{c}{(----------------------- in ''/cy ---------------------)} & \\
\noalign{\smallskip}
\hline 
\noalign{\smallskip}
Lieske et al., 1977         &                  &  5038.7784        & {\bf 5029.0966} & -1.92     & $23^{\circ}26'21''.448$  \\
\noalign{\smallskip}
Kinoshita, 1977 and Seidelmann, 1982& 3.2739935&  5038.7784        & {\bf 5029.0966} & -1.92     & $23^{\circ}26'21''.448$  \\
\noalign{\smallskip}
\hline
\noalign{\smallskip}
Williams, 1994              &  3.2737634       & {\bf 5038.456501} &  5028.7700      & -1.9194   & $23^{\circ}26'21''.409$  \\
\noalign{\smallskip}
Souchay \&  Kinoshita, 1996 &  3.2737548       &  ~~~~~~~-         &  5028.7700      & -1.9194   & $23^{\circ}26'21''.448$  \\
\noalign{\smallskip}
Bretagnon et al., 1997      &  3.2737671       & 5038.456488       &{\bf 5028.7700}  & -1.919883 & $23^{\circ}26'21''.412$  \\
\noalign{\smallskip}
Bretagnon et al., 2003      &  ~~~~~~-         & 5038.478750       &{\bf 5028.792262}& -1.919883 & $23^{\circ}26'21''.40880$\\
\noalign{\smallskip}
Fukushima, 2003             &  3.2737804       & {\bf 5038.478143} &  5028.7955      & -1.9196	 & $23^{\circ}26'21''.40955$\\
\noalign{\smallskip}
Capitaine et al., 2003      &  3.27379448      & {\bf 5038.481507} &  5028.796195    & -1.919883 & $23^{\circ}26'21''.406$  \\
\noalign{\smallskip}
\hline
\noalign{\smallskip}
Mathews et al., 2002        &  3.27379492      & {\bf 5038.478750} &  5028.7923      & -1.9198	 & $23^{\circ}26'21''.410$  \\
\noalign{\smallskip}
\hline
\end{tabular}
\end{table*}

The computation of the IAU~2000 precession-nutation model by Mathews et al.~(2002) is based on a new method which uses geophysical considerations. They adjust nine Basic Earth Parameters (BEP), including the Earth dynamical flattening $H$.


\subsection{Method and parameters used in this study}

Based on the paper by Capitaine et al. (2003), denoted hereafter P03, we use differential equations (\ref{eq:Syst_Prec_Nut}) in which $H$ has been replaced by $H+\Delta H$ (using Eq.~(\ref{eq:luni_solar_effect}), Eq.~(\ref{eq:Km}), Eq.~(\ref{eq:Ks}) and Eq.~(\ref{eq:r0})). We start from the P03 initial values for the variables $\omega_A$, $\psi_A$, $\epsilon_A$, $\chi_A$ and $p_A$, that are represented as polynomials of time and rely on the numerical values given in Table \ref{tab:our_model}. We solve Eq.~(\ref{eq:Syst_Prec_Nut}) together with the other precession equations (e.g. see Eq.~(26) and Eq.~(28) of P03) with the software GREGOIRE (Chapront, 2003) that can process Fourier and Poisson expressions. We iterate this process until we obtain a convergence of the solution.

\begin{table}[htbp]
\centering
\caption{Numerical values used in this study. $H$, $\psi_1$ and $\omega_1$ are integration constants.}\label{tab:our_model}
\begin{tabular}{cl}
\hline
\hline
\noalign{\smallskip}
\multicolumn{2}{c}{Initial values at J2000.0}           \\
\noalign{\smallskip}
\hline
\noalign{\smallskip}
$H$                             & $H_{\mbox{\scriptsize MHB}}=3.27379492 \times 10^{-3}$ \\   
\noalign{\smallskip}
$\psi_1$                        & $5038''.481507$/cy                  \\
\noalign{\smallskip}
$\omega_1$                      & $-0''.02575$/cy                     \\
\noalign{\smallskip}
\noalign{\smallskip}
\noalign{\smallskip}
$p_1$                           & $5028''.796195$/cy                  \\
\noalign{\smallskip}
$\chi_1$                        & $10''.556403$/cy                    \\
\noalign{\smallskip}
$\epsilon_0$                    & $84381''.406 = 23^{\circ} 26' 21''.406$  \\
\noalign{\smallskip}
\hline
\noalign{\smallskip}
\multicolumn{2}{c}{}  \\
\noalign{\smallskip}
\hline
\hline
\noalign{\smallskip}
\multicolumn{2}{c}{Contributions to the precession rate in longitude (in $''$/cy)}   \\
\noalign{\smallskip}
\hline
\noalign{\smallskip}
Lunisolar first order           & $~~5494.062986 \times \cos \epsilon_0 \simeq 5040.7047$   \\
Planetary first order           & $~~0.031$                  \\ 
Geodesic precession             & $-1.919882$              \\
\noalign{\smallskip}
\hline
\end{tabular}
\end{table}


\section{Relationship between $C_{20}$ and $H$}

\subsection{Relation}

From the geodetic $C_{20}$ variation series we can derive the corresponding variations of the dynamical flattening $H$. Indeed, knowing that $J_2=-~C_{20}=-~\sqrt{5}~\bar{C}_{20}$, in the case of a rigid Earth, we can write (see Lambeck, 1988):
\begin{eqnarray}\label{eq:Hd}
H = \left(C-\frac{A+B}{2}\right) /C & = & \frac{M ~{R_e}^2}{C} ~J_2                                \\ 
                                    & = & - \frac{M ~{R_e}^2}{C} ~C_{20}                 \nonumber \\ 
                                    & = & - \sqrt{5} ~\frac{M ~{R_e}^2}{C} ~\bar{C}_{20} \nonumber 
\end{eqnarray}
where $A$, $B$ and $C$ are the mean equatorial and polar moments of inertia of the Earth. $M$ and $R_e$ are respectively the mass and the mean equatorial radius of the Earth. $\bar{C}_{20}$ is the normalized Stokes coefficient (of degree 2 and order 0) of the geopotential. 

But the Earth is elastic, so let us consider small variations of $H$, $C_{20}$ and the third principal moment of inertia of the Earth ($C$ being its constant part and $c_{33}$ its variable part). Then we obtain:
\begin{equation}\label{eq:H_total}
H_{\mbox{\scriptsize ~total}} = \frac{M ~{R_e}^2}{C} ~\frac{1}{1+\frac{c_{33}}{C}} ~{J_2}_{\mbox{\scriptsize ~total}}     
\end{equation}
$c_{33}/C$ being a small quantity of the order of $10^{-6}$, we consider the Taylor development of $\left( 1+c_{33}/C \right)^{-1}$. Then the total expression of $H$ can be written as:
\begin{equation}\label{eq:H_total_bis}
H_{\mbox{\scriptsize ~total}} = \frac{M {R_e}^2}{C} ~{J_2}_{\mbox{\scriptsize ~total}} \left( 1 - \frac{c_{33}}{C} + \left( \frac{c_{33}}{C} \right)^2 + ... \right)
\end{equation}
where $M {R_e}^2/C \times \left( c_{33}/C \right)^n J_2$ for $n \geq 1$ is smaller than $10^{-11}$. So, in Eq.~(\ref{eq:H_total_bis}), considering (i) constant and variable parts separately and (ii) Eq.~(\ref{eq:Hd}), we obtain:
\begin{equation}\label{eq:Delta_Hd}
\Delta H = \frac{M {R_e}^2}{C} ~\Delta J_2 = - \sqrt{5} ~\frac{M {R_e}^2}{C} ~\Delta \bar{C}_{20}
\end{equation}
where $\Delta J_2 = - \Delta C_{20} = - \sqrt{5} ~\Delta \bar{C}_{20}$ corresponds to the variations of the Stokes coefficient $J_2$. Generally, we can write: $\Delta J_2 \propto c_{33}/C$ (Lambeck, 1988).


\subsection{Computation of the ratio $M {R_e}^2/C$}

The coefficient $M {R_e}^2/C$ is usually obtained from the $H$ and $J_2$ values (see Eq.~(\ref{eq:Hd})). In order to determine the constant part of $H$, we can use (i) the $R_e$, $M$ and $C$ values or (ii) the Clairaut theory (see Table \ref{tab:Coeff_et_H}). 

First, recall the Earth geometrical flattening $\epsilon$:
\begin{eqnarray}\label{eq:eps} 
\epsilon = \frac{R_e - R_p}{R_e} 
\end{eqnarray}
where $R_p$ and $R_e$ are respectively the polar and equatorial mean radii of the Earth.
Second, recall the assumptions that the Earth (i) is in hydrostatic equilibrium and (ii) is considered as a revolutional ellipsoid. Hence, the first Clairaut equation gives the Earth geometrical flattening as a function of $J_2$ and $q$. The approximations to the first and second order are respectively:
\begin{eqnarray}
\epsilon & = & \frac{q}{2} + \frac{3}{2} J_2  \label{eq:eps_via_q_J2} \\
\epsilon & = & \frac{q}{2} + \frac{3}{2} J_2 + \frac{9}{8} {J_2}^2 - \frac{3}{14} J_2~q - \frac{11}{56} q^2 \label{eq:eps_via_q_J2_bis} 
\end{eqnarray}
where the geodynamical constant is:
\begin{eqnarray}\label{eq:q}
q & = & \frac{\omega^2 {R_e}^3}{GM}    \\
  & = & 3.461391 \times 10^{-3} ~~\mbox{, IAG (Groten, 1999).} \nonumber
\end{eqnarray}
Then, the following Radau equation can help us to determine the expression of $M {R_e}^2/C$:
\begin{equation}\label{eq:Radau}
\frac{\epsilon - q/2}{H} = 1-\frac{2}{5} \sqrt{1+\eta} = \frac{1}{\lambda}
\end{equation}
where $\lambda$ is the d'Alembert parameter and $\eta$ the Radau parameter, as:
\begin{equation}\label{eq:eta}
\eta = \frac{5 q}{2 \epsilon} - 2
\end{equation}
Hence, replacing $\epsilon$ with Eq.~(\ref{eq:eps_via_q_J2}) in Eq.~(\ref{eq:Radau}) and using Eq.~(\ref{eq:Hd}) gives the Darwin-Radau relation as following:
\begin{equation}\label{eq:CMRe2}
\frac{C}{M {R_e}^2} = \frac{2}{3 \lambda} = \frac{2}{3} \left( 1-\frac{2}{5} \sqrt{1+\eta} \right) 
\end{equation}
Our tests have shown that Eq.~(\ref{eq:eps_via_q_J2_bis}) for $\epsilon$, in the expression (\ref{eq:eta}) of $\eta$, gives more reliable results.

In Table \ref{tab:Coeff_et_H} we compare the various $H$ values obtained. We denote (i) $H^*$ the value obtained with the Clairaut method and (ii) $H^{**}$ the value obtained using directly the $R_e$, $C$ and $M$ values. Both are computed with Eq.~(\ref{eq:Hd}) and a value for $J_2$ of $1.0826358 \times 10^{-3}$. Note that in contrast, IAG or MHB values (usually used) are determined from astronomical precession observations and can be used to compute the $C/M{R_e}^2$ value. We can add that the differences with $H_{MHB}$ come from (i) the hydrostatic equilibrium hypothesis in Clairaut's theory for the value $H^*$ and (ii) the poorly determined $R_e$, $C$ and $M$ values, for the value $H^{**}$. This will introduce errors in the $\Delta H$ determination, which we will study in Sect.~3.3. 

In the following, we will use the $C/\left(M {R_e}^2\right)$ value determined with the Clairaut theory, noted with a (*) in Table \ref{tab:Coeff_et_H}, which corresponds to a value for $H$ of: $H^* = 3.26715240 \times 10^{-3}$.

\begin{table*}[htbp]
\centering
\caption{Comparison between different values of the coefficient $C/\left(M {R_e}^2\right)$ and of the constant part for $H$: (1) IAG values (Groten, 1999) - (2) MHB values (Mathews et al., 2002) - (3) Constant part $H^{**}$ obtained from Eq.~(\ref{eq:Hd}) using the $M$, $R_e$ and $C$ IAG values - (4) Method of ``Clairaut'' (Sect. 3.2), assuming hydrostatic equilibrium. The third and fourth methods use a constant part for $\bar{C}_{20}$ of $-4.841695 \times 10^{-4}$ in Eq.~(\ref{eq:Hd}) (i.e. $J_2=1.0826358 \times 10^{-3}$). The sense of the computation is indicated by the arrows.}\label{tab:Coeff_et_H}
\begin{tabular}{lcccc}
\hline
\hline
\noalign{\smallskip}
			   &       (1)           &          (2)         &          (3)                &       (4)      \\
                           &   IAG (1999)        &       MHB~2000       &  Separate values for        &    Clairaut    \\
			   &                     &                      &  $M$, $R_e$ and $C$         &    Theory      \\
\noalign{\smallskip}
\hline
\noalign{\smallskip}
$C/\left(M~{R_e}^2\right)$ &	0.330701         &   0.330698           &  $0.330722^{**}$            &  $0.331370^*$   \\
                           & $\pm 2 \times 10^{-6}$    & \multicolumn{3}{c}{} \\
\noalign{\smallskip}
			   &    $\Uparrow$       &    $\Uparrow$        &   $\Downarrow$              &   $\Downarrow$	\\
\noalign{\smallskip}
$H$                        & $3.273763 \times 10^{-3}$ & $3.27379492 \times 10^{-3}$ & $H^{**}=3.27355562 \times 10^{-3}$ & $H^*=3.26715240 \times 10^{-3}$\\
                           & $\pm 2 \times 10^{-8}$    & \multicolumn{3}{c}{} \\
\noalign{\smallskip}
\hline
\end{tabular}
\end{table*}


\subsection{Error estimation}

We can estimate the error that the use of the Clairaut theory introduces into the $\Delta H$ results. Indeed, if we consider the MHB value as the realistic $H$ value (see Table \ref{tab:Coeff_et_H}), the relative error made is:
\begin{equation}\label{eq:relat_error}
\sigma_H = \frac{H_{\mbox{\scriptsize MHB}} - H^*}{H_{\mbox{\scriptsize MHB}}} \simeq 2 \times 10^{-3}
\end{equation}
We estimate that the error is about 0.2 \%. So, computing the variable part of $H$ with the $C_{20}$ data results in a maximum error of about:
\begin{eqnarray}\label{eq:relat_error_variable}
| {\Delta H}_{\mbox{\scriptsize real}} - {\Delta H^*} | & \simeq & \left(2\times10^{-3}\right) \times \left(6\times10^{-9}\right) \nonumber\\
                                     & \simeq & 1.2 \times 10^{-11}
\end{eqnarray}
assuming that the maximum value for $\Delta H$ is of the order of $6 \times 10^{-9}$. Then, regarding the values of the $\Delta H$ data and of their precision, we can consider this error as negligible.


\section{Time series of $\Delta C_{20}$ used in this study}

\begin{figure}
\centering
\includegraphics[width=6cm, angle=-90]{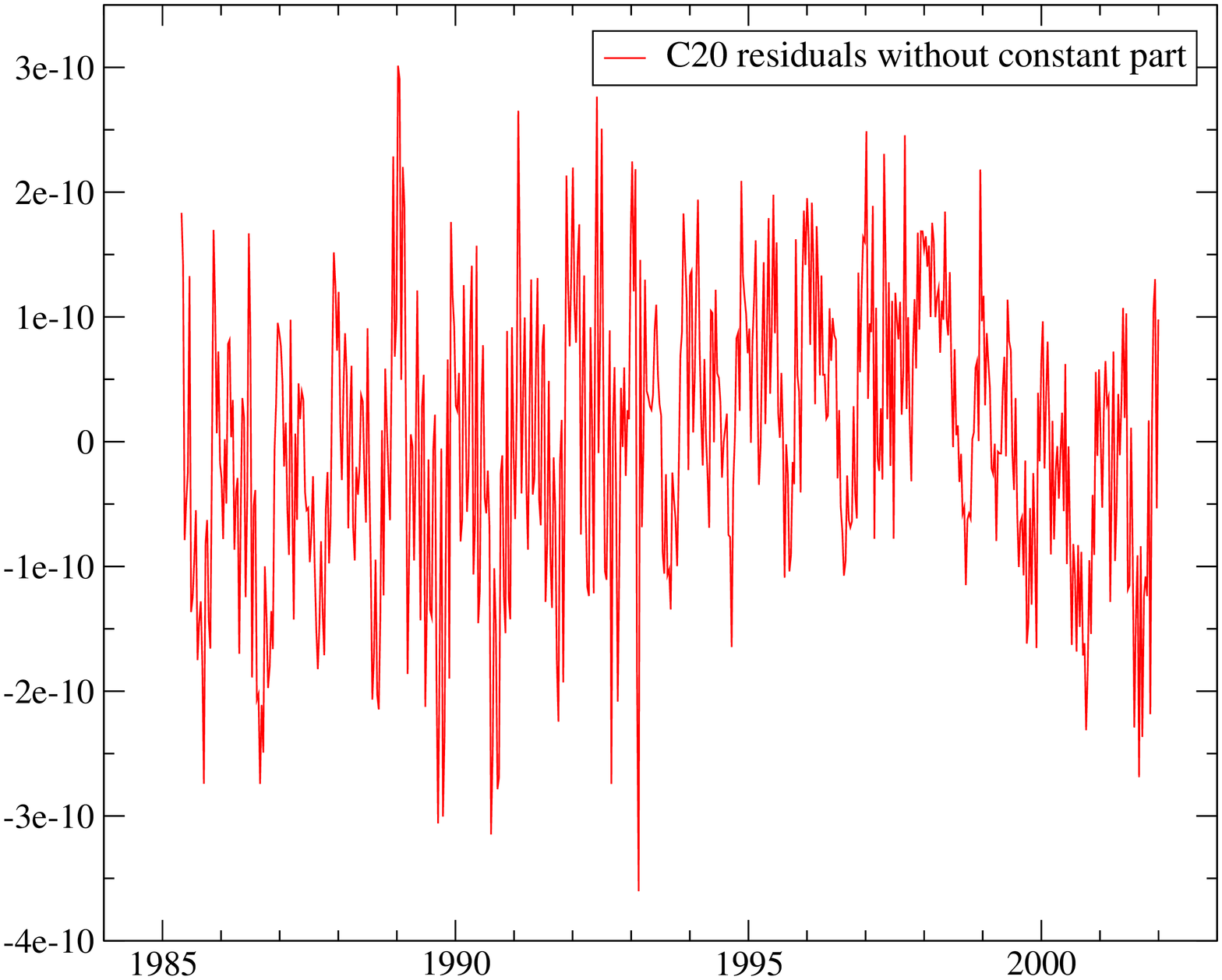}
\includegraphics[width=6cm, angle=-90]{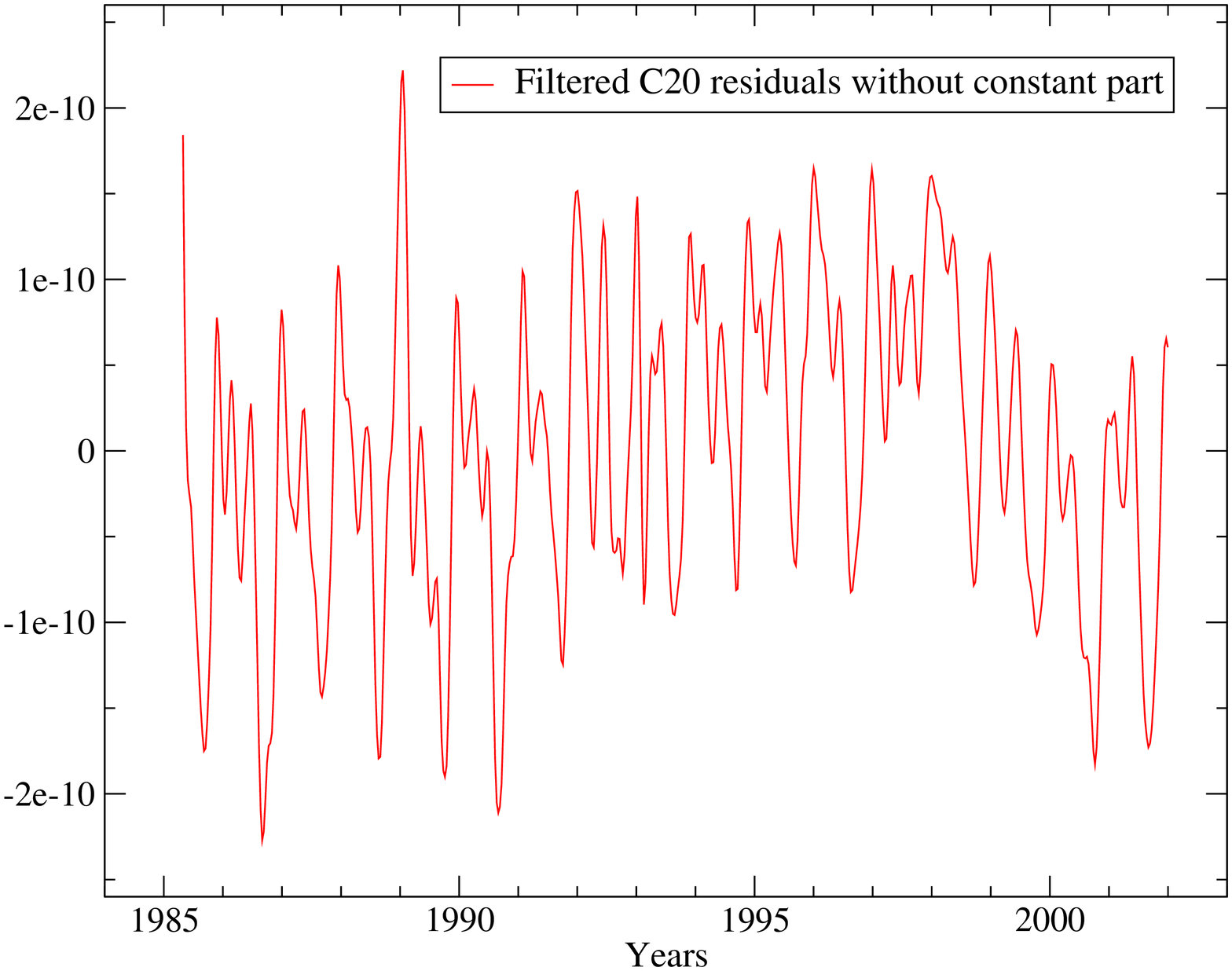}
\caption{Normalized $\Delta C_{20}$ residuals (top: raw residuals, bottom: filtered residuals, where the high frequency signals have been removed): non-modelled part of the $\Delta C_{20}$ harmonic coefficient of the Earth gravity field.}
\label{fig:C20_grgs}
\end{figure}

The geodetic data used are the time series (variable part) of the spherical harmonic coefficient $C_{20}$ of the geopotential, obtained by the GRGS (Groupe de Recherche en G\'{e}od\'{e}sie Spatiale, Toulouse) from the precise orbit determination of several satellites (like LAGEOS, Starlette or CHAMP) from 1985 to 2002 (Biancale et al., 2002). The combination of these satellites allows the separation of the different zonal geopotential coefficients, more particularly of $J_2$ and $J_4$.
This series includes (i) a model part for the atmospheric mass redistributions (Chao \& Au, 1991; Gegout \& Cazenave, 1993) and for the oceanic and solid Earth tides (McCarthy, 1996), and (ii) a residual part (see Fig.~\ref{fig:C20_grgs}) obtained as difference of the space measurements with respect to a model. These various changes in the Earth system are modelled as variations in the standard geopotential coefficient $C_{20}$ and we note the different contributions $\Delta C_{{20}_{atm}}$, $\Delta C_{{20}_{oc}}$, $\Delta C_{{20}_{soltid}}$ and $\Delta C_{{20}_{res}}$, respectively.

\subsection{$\Delta C_{20}$ residuals and its secular trend: observed part}

Earlier studies already took into account the effect of the secular variation of $C_{20}$ on the precession of the equator. Such a secular variation is attributed to the post-glacial rebound of the Earth (Yoder et al., 1983), which reduces its flattening. Williams (1994) and Capitaine et al. (2003) considered a $J_2$ rate value of $-3\times10^{-9}$/cy. Using the numerical value of Table \ref{tab:our_model} for the first order contribution ($f_{01} \cos \epsilon_0$) to the precession rate $r_0$, which is directly proportional to $J_2$, the contribution $\dot{J_2}/J_2 \times f_{01} \cos \epsilon_0$ of the $J_2$ rate to the acceleration of precession $d^2\psi_A/dt^2$ is about $-0.014 ''$/cy$^2$, giving rise to a $-0.007 ''$/cy$^2$ contribution to the $t^2$ term in the expression of $\psi_A$.

\begin{figure}
\centering
\includegraphics[width=6cm, angle=-90]{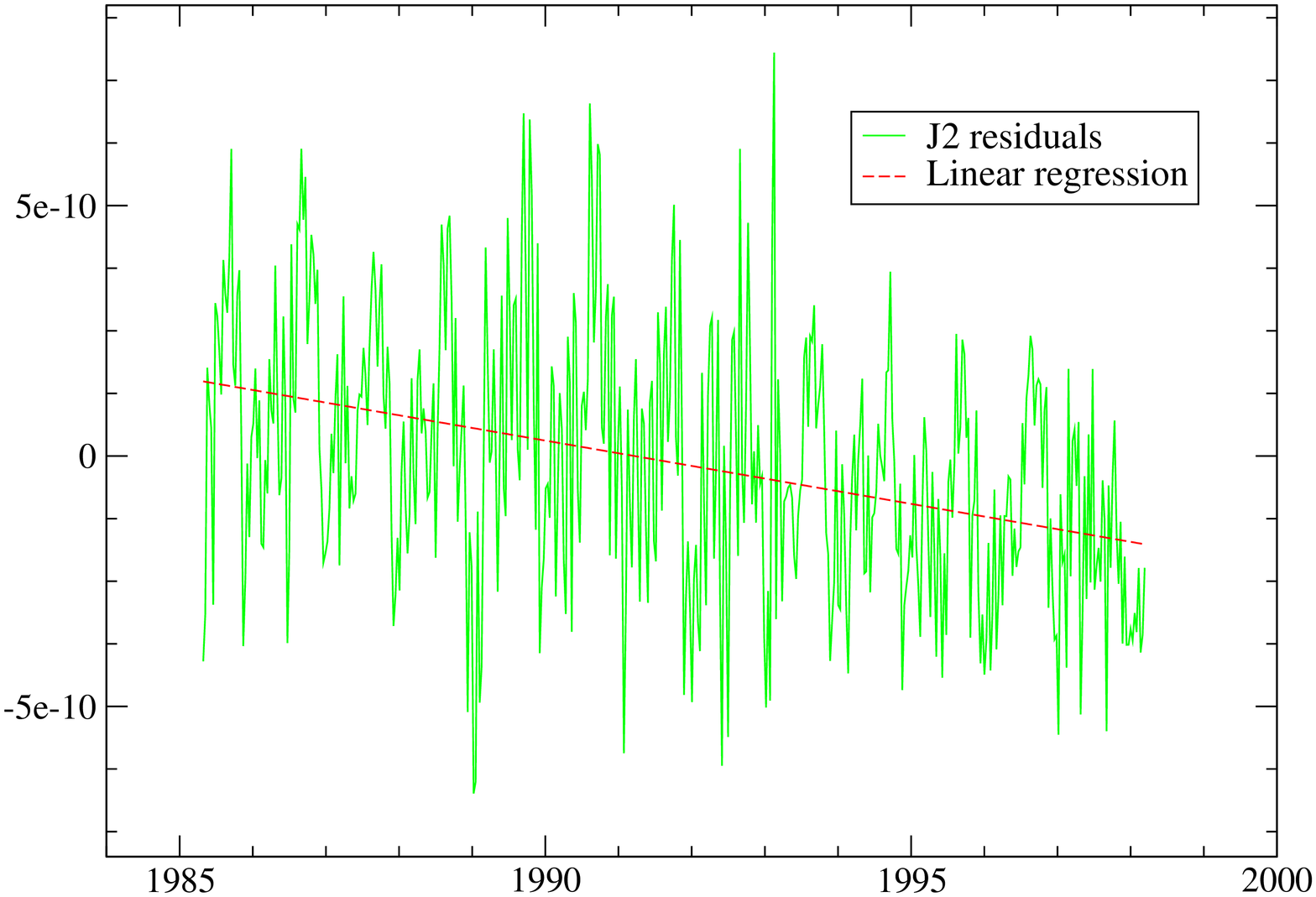}
\includegraphics[width=6cm, angle=-90]{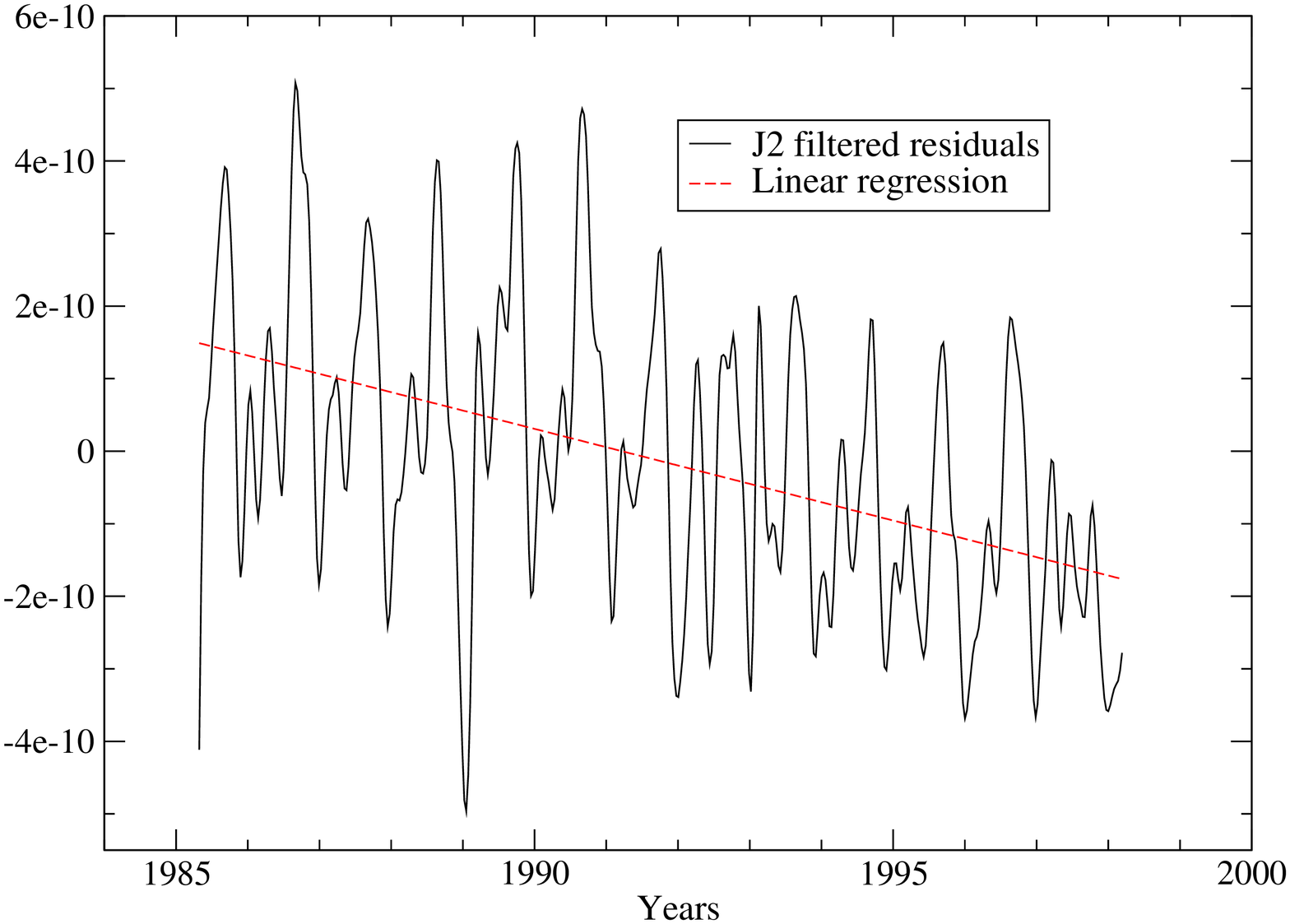}
\caption{$J_2$ GRGS residuals (top: raw residuals, bottom: filtered residuals, where the high frequency signals have been removed): estimation of the linear trend, from 1985 to 1998.}
\label{fig:J2_residuals}
\end{figure}

Since $1998$, a change in the secular trend of the $J_2$ data has been reported (Cox \& Chao, 2000). This change can be seen in the series of $\Delta \bar{C}_{20}$ residuals (see Fig.~\ref{fig:C20_grgs}). An attempt to model this effect, with oceanic data, water coverage data and geophysical models, has been investigated by Dickey et al.~(2002). Using the residuals $\Delta \bar{C}_{20}$ of the GRGS, we can estimate a secular trend for $J_2=-\sqrt{5}~\bar{C}_{20}$ from 1985 to 1998 (see Fig.~\ref{fig:J2_residuals}). We find a $J_2$ rate of the order of: $-2.5 ~(\pm 0.2) \times 10^{-9}$/cy, which gives a change of about $-0.006$''/cy$^2$ in the $t^2$ term of the polynomial development of the precession angle $\psi_A$. 

As this secular trend is not the same in the total data span, we will also model the long term variations in the $C_{20}$ residual series with a periodic signal. Such a long-period term in the $J_2$ residual series may come from mismodelled effects, particularly from the $18.6$-yr solid Earth tides. We will make such an assumption and adjust for the period 1985-2002, a secular trend and a long-period term in the $\Delta C_{20}$ residual series (see Sect. 4.3). 

However, it should be noted that a secular trend for $J_2$, of the order of $-3 \times 10^{-9}$/cy, is more consistent with long term studies of the Earth rotation variations by Morrison \& Stephenson (1997), based upon eclipse data over two millennia (they found $\dot{J}_2= \left( -3.4\pm0.6 \right) \times 10^{-9}$/cy).

\subsection{$\Delta C_{20}$ geophysical data used: modelled part}

The geophysical models that have been previously subtracted from the $C_{20}$ data (i.e. atmospheric, oceanic and solid Earth tides effects) must be added back to these data in exactly the same way they had been subtracted to reconstruct the relevant geophysical contributions.

For each contribution we give the associated potential $U$ at the point $(r, \phi, \lambda, t)$ (limited to the degree 2 and order 0) that we identify with the Earth gravitational potential. Hence, we obtain the $\Delta \bar{C}_{20}$ coefficient contribution of each geophysical source. \\

\begin{figure}
\centering
\includegraphics[width=6cm, angle=-90]{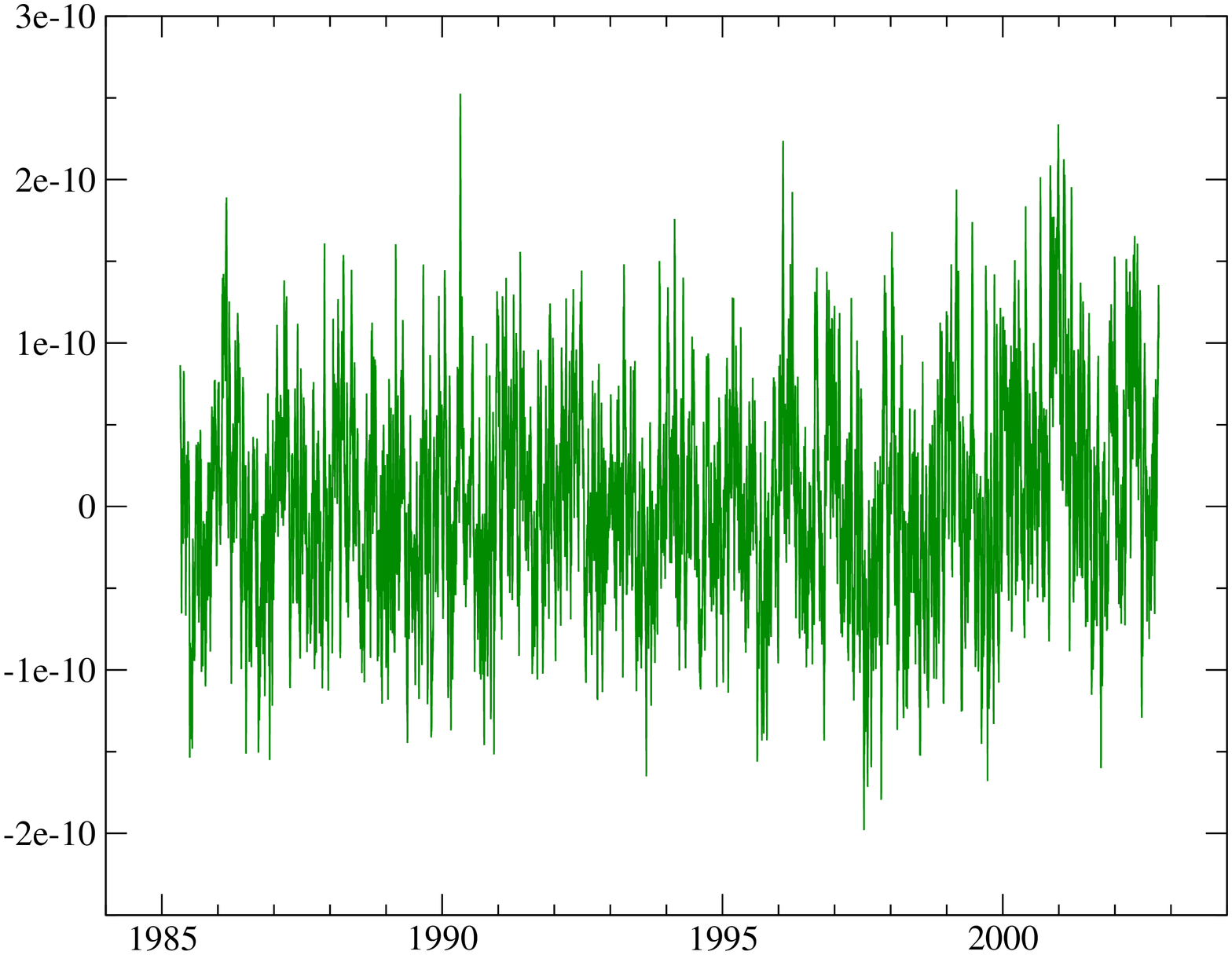}
\includegraphics[width=6cm, angle=-90]{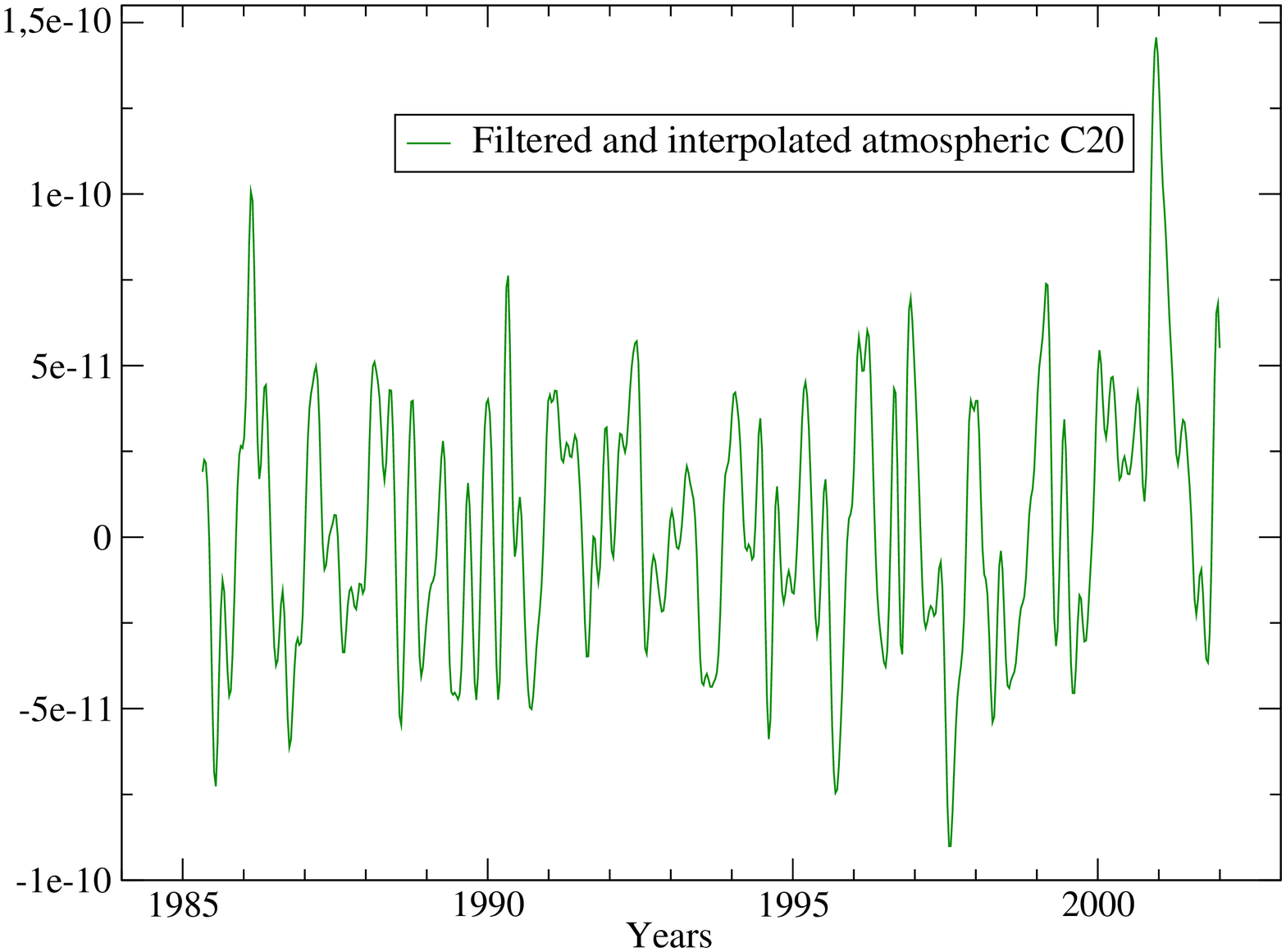}
\caption{Normalized atmospheric $\Delta C_{20}$ (top: raw data, bottom: filtered data, where the high frequency signals have been removed): atmospheric modelled part of the $\Delta C_{20}$ harmonic coefficient of the Earth gravity field, obtained with ECMWF pressure data.}
\label{fig:C20_atm}
\end{figure}

$\bullet$ The atmospheric contribution is due to pressure changes in time, measured and given by the European Centre for Medium-range Weather Forecasts (ECMWF) (see Fig.~\ref{fig:C20_atm}). The simple-layer atmospheric potential, limited to degree 2 and order 0, can be expressed as:
\begin{equation}\label{eq:pot_atm}
\begin{array}{l}
U_{\mbox{\scriptsize atm}} =  4 \pi G R_e ~\frac{1+k_2^{'}}{5 g} \left( \frac{R_e}{r} \right)^{3} \Delta \bar{C}_{{20}_{\mbox{\scriptsize ECMWF}}}(t) ~\bar{P}_{20}(\sin \phi)
\end{array}
\end{equation}
where $G=6.672 \times 10^{-11}$~m$^3$~kg$^{-1}$~s$^{-2}$ is the gravitational constant, $k_2^{'}=-0.314166$ is a Love number (Farrell, 1972), $g=9.81$~m~s$^{-2}$, and  $\bar{P}_{20}(\sin \phi)$ is the Legendre function of degree 2 and order 0. The $\bar{C}_{{20}_{\mbox{\scriptsize ECMWF}}} (t)$ atmospheric coefficient, expressed in Pascals, comes from the spherical harmonic decomposition of the ECMWF atmospheric pressure grids, every 6 hours, over continents (see Gegout \& Cazenave (1993) or Chao \& Au (1991)):
\begin{equation}
\Delta \bar{C}_{{20}_{\mbox{\scriptsize ECMWF}}} (t) = \int_{S} \Delta p~(\phi, \lambda, t)~\left(\frac{3}{2}\sin^2\phi-\frac{1}{2} \right)~dS
\end{equation}
where $S$ is a surface grid pressure around the Earth and $\Delta p$ is the difference of pressure with a constant part prefixed, at the point $(\phi, \lambda)$.
Hence, identifying Eq.~(\ref{eq:pot_atm}) with the Earth gravitational potential, we obtain the atmospheric pressure contribution to the $\Delta \bar{C}_{20}$ harmonic coefficient: 
\begin{equation}
\Delta \bar{C}_{{20}_{\mbox{\scriptsize atm}}} (t) = \frac{4 \pi ~ {R_e}^2 ~(1+k_2^{'})}{5 M g} ~\Delta \bar{C}_{{20}_{\mbox{\scriptsize ECMWF}}} (t) 
\end{equation}

\vspace{0.5cm}

\begin{figure}
\centering
\includegraphics[width=6cm, angle=-90]{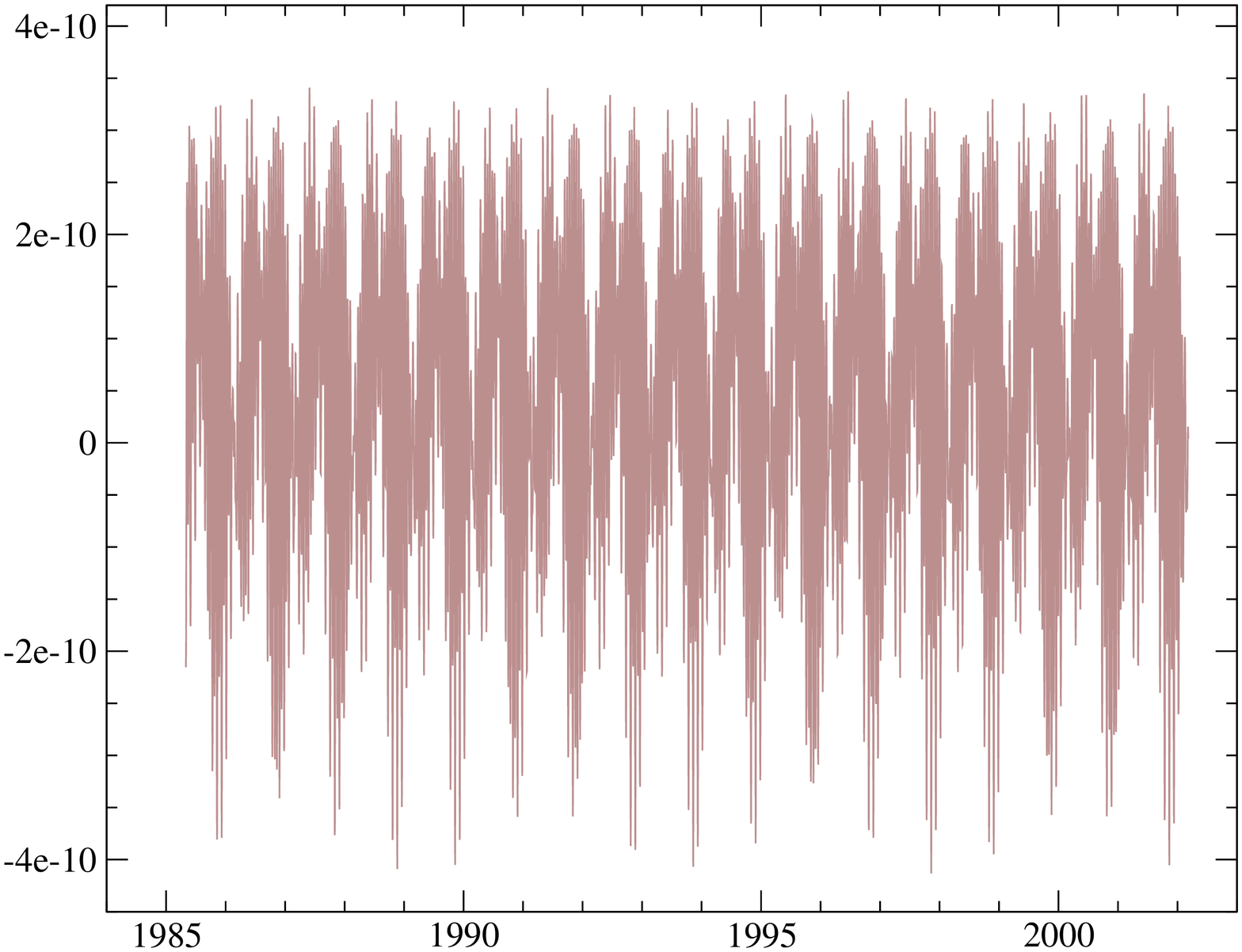}
\includegraphics[width=6cm, angle=-90]{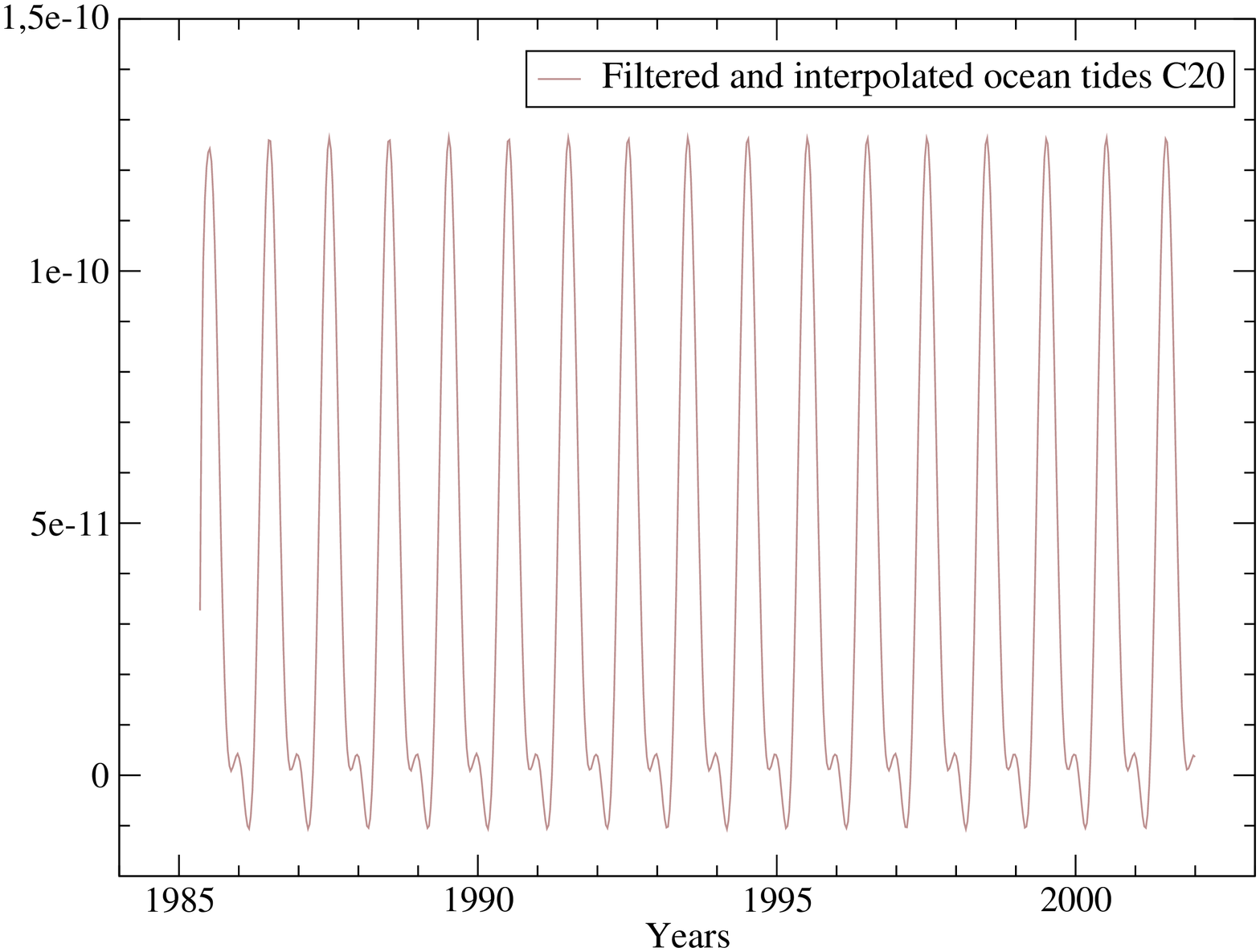}
\caption{Normalized oceanic $\Delta C_{20}$ (top: raw data, bottom: filtered data, where the high frequency signals have been removed): oceanic-tide-modelled part of the $\Delta C_{20}$ harmonic coefficient of the Earth gravity field; IERS Conventions 1996.}
\label{fig:C20_ocean_tide}
\end{figure}

$\bullet$ The contribution of the oceanic tides (see Fig.~\ref{fig:C20_ocean_tide}) is modelled in the IERS Conventions 1996. The Earth responds to the dynamical effects of ocean tides, and the associated potential, limited to the degree 2 and order 0, is:
\begin{equation}\label{eq:pot_oc_tide}
U_{\mbox{\scriptsize oc}} = 4 \pi ~G ~R_e ~\rho_w ~\frac{1+k_2^{'}}{5} \left( \frac{R_e}{r} \right)^{3} ~\bar{P}_{20}(\sin \phi) ~\alpha (t)
\end{equation}
where we note $\alpha$, depending on time, as:
\begin{equation}
\alpha = \sum_n \sum_{+}^{-} C_{n,2,0}^{\pm} ~\cos(\theta_n(t) + \chi_n) + S_{n,2,0}^{\pm} ~\sin(\theta_n(t) + \chi_n) 
\end{equation}
The sum over $n$ corresponds to the Doodson development whose associated arguments are $\theta_n$ and $\chi_n$. The parameter $\rho_w$ ($= 1025$~kg~m$^{-3}$) is the mean density of sea water. Furthermore, $ C_{n,2,0}^{\pm} = \hat{C}_{n,2,0}^{\pm} ~\sin(\epsilon_{n,2,0}^{\pm})$ and  $S_{n,2,0}^{\pm} =  \hat{C}_{n,2,0}^{\pm} ~\cos(\epsilon_{n,2,0}^{\pm})$, where $\hat{C}_{n,2,0}^{\pm}$ and $\epsilon_{n,2,0}^{\pm}$ are the normalized amplitude and phase of the harmonic model of the oceanic tides limited to degree 2 and order 0. Identifying Eq.~(\ref{eq:pot_oc_tide}) with the Earth gravitational potential gives the oceanic tide contribution to the $\Delta \bar{C}_{20}$ harmonic coefficient: 
\begin{equation}
\Delta \bar{C}_{{20}_{\mbox{\scriptsize oc}}} (t) = \frac{4 \pi ~{R_e}^2 ~(1+k_2') ~\rho_W}{5 ~M} ~\alpha (t) 
\end{equation}

\vspace{0.5cm}

\begin{figure}
\centering
\includegraphics[width=6cm, angle=-90]{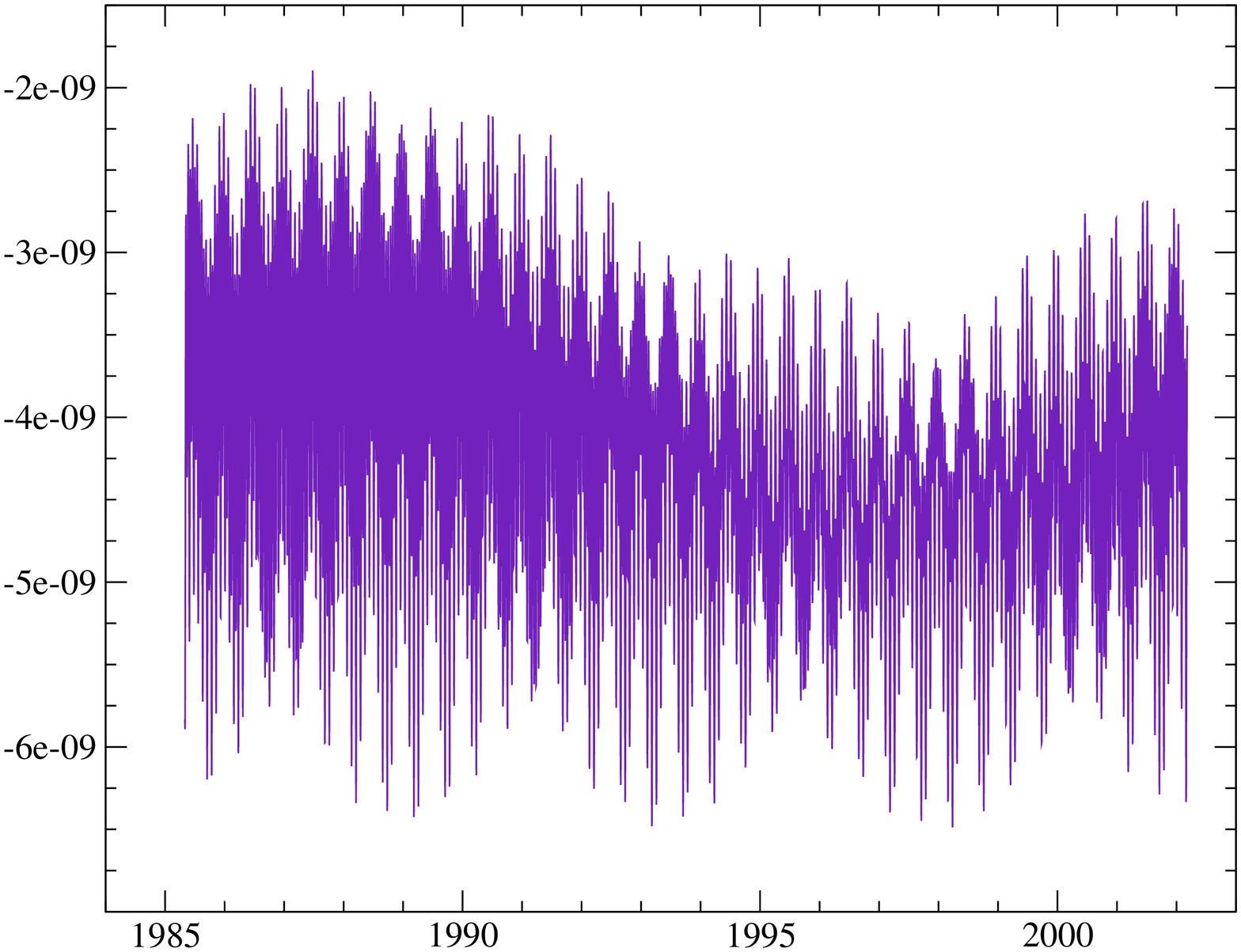}
\includegraphics[width=6cm, angle=-90]{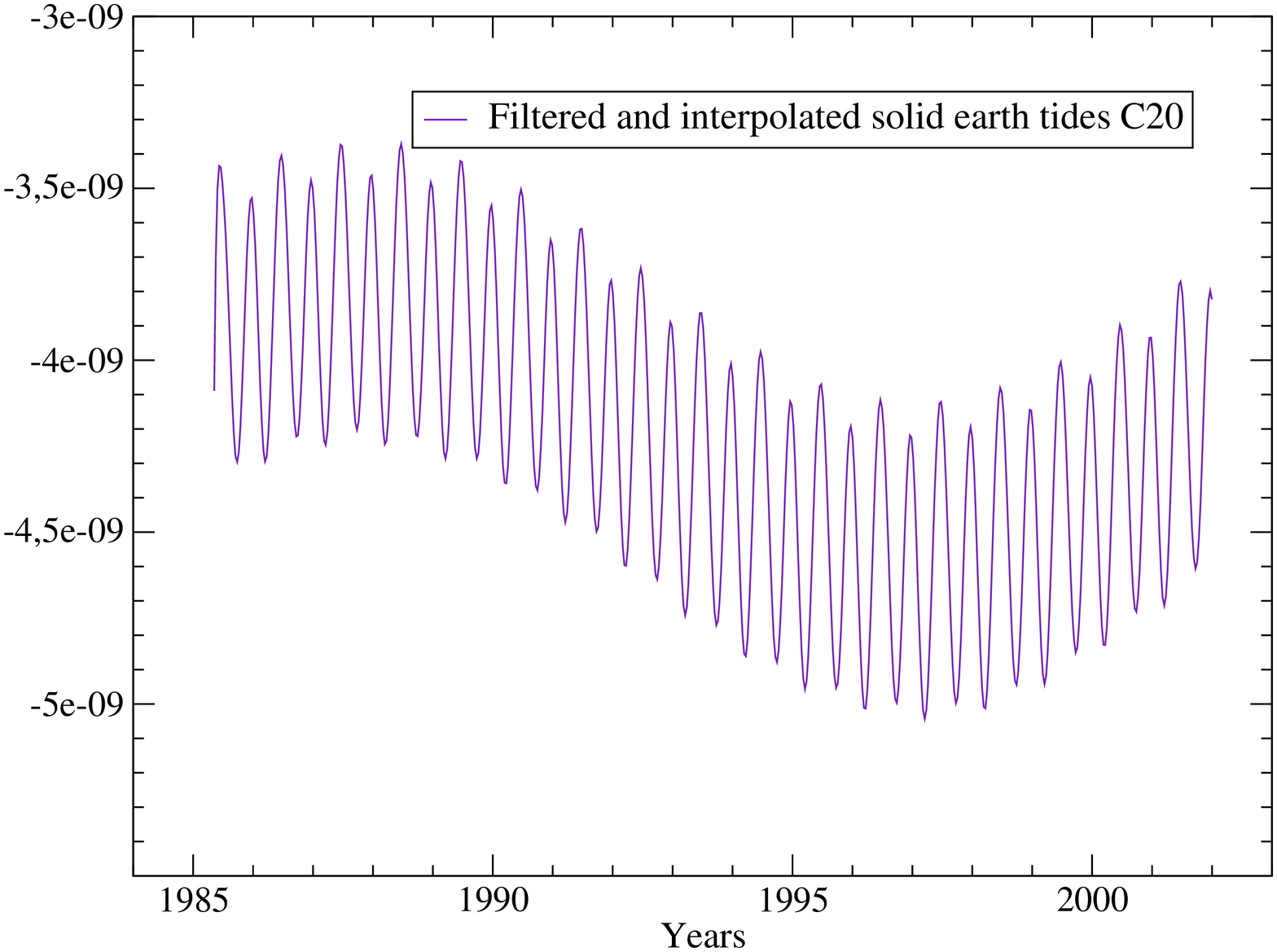}
\caption{Normalized solid tides $\Delta C_{20}$ (top: raw data, bottom: filtered data, where the high frequency signals have been removed): solid-Earth-tide-modelled part of the $\Delta C_{20}$ harmonic coefficient of the Earth gravity field; IERS Conventions 1996.}
\label{fig:C20_soltid}
\end{figure}

$\bullet$ The solid Earth tide contribution (see Fig.~\ref{fig:C20_soltid}) is due to the gravitational effect of the Moon and the Sun on the Earth (IERS Conventions 1996). This force derives from a potential, developed in spherical harmonics, which limited to degree 2 and order 0 is:
\begin{equation}\label{eq:pot_sol_tide}
U_{\mbox{\scriptsize soltid}} = G ~M ~\frac{{R_e}^2}{r^3} ~\bar{P}_{20} (\sin \phi) ~\bar{C}_{{20}_{\mbox{\scriptsize Moon+Sun}}}(t) 
\end{equation}
where 
\begin{equation}
\bar{C}_{{20}_{\mbox{\scriptsize Moon+Sun}}} (t) = \frac{k_{20} ~{R_e}^3}{5 ~M} \sum_{p=moon}^{sun} \left( \frac{m_{p}}{r_{p}^3} ~\bar{P}_{20} (\sin \phi_{p}) \right)
\end{equation}
where $k_{20}=0.3019$ is the nominal degree Love number for degree 2 and order 0, $m_p$ the mass of the body $p$, and $r_p$ the geocentric distance and $\phi_p$ the geocentric latitude at each moment of the body $p$. The Love number depends on the tidal frequencies acting on the Earth. Hence, the contribution to $\Delta \bar{C}_{20}$ from the long period tidal constituents of various frequencies $\nu$ must be corrected (see IERS Conventions 1996). 
Eq.~(\ref{eq:pot_sol_tide}) corrected for the frequency dependence of the Love number, can be identified with the Earth gravitational potential. We obtain the Earth solid tide contribution to the $\Delta \bar{C}_{20}$ harmonic coefficient: 
\begin{equation}
\Delta \bar{C}_{{20}_{\mbox{\scriptsize soltid}}} (t) = \bar{C}_{{20}_{\mbox{\scriptsize Moon+Sun}}} + ``frequency ~correction"
\end{equation}
This contribution comprises a constant part in the $\Delta \bar{C}_{20}$ solid Earth tide, which is called ``permanent tide''. We have estimated it and obtained: $-4.215114 \times 10^{-9}$ (the IERS Conventions value is $-4.201 \times 10^{-9}$). We must remove it from our $\Delta \bar{C}_{20}$ data coming from solid Earth tides.

\vspace{0.5cm}

\begin{figure}
\centering
\includegraphics[width=6cm, angle=-90]{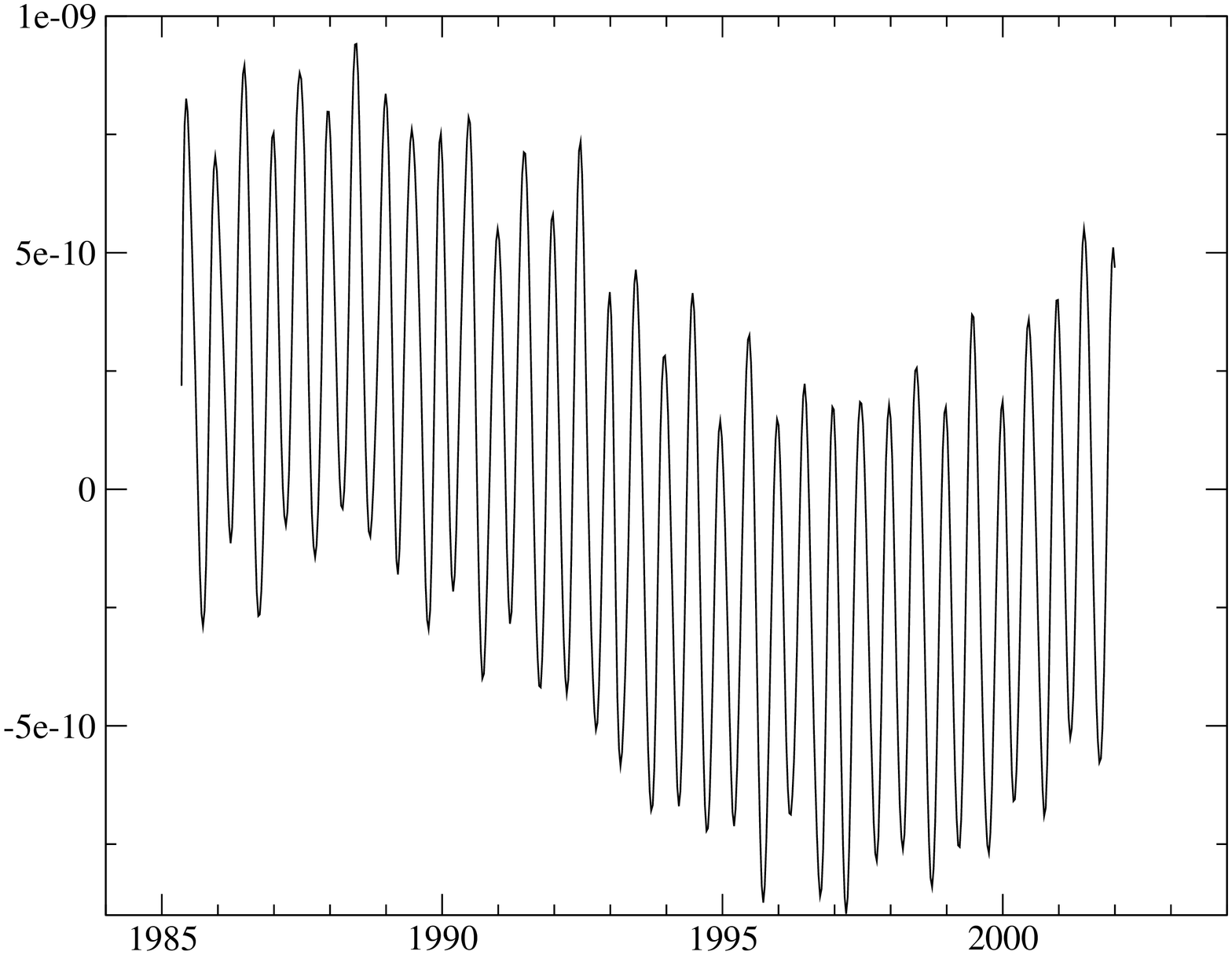}
\includegraphics[width=6cm, angle=-90]{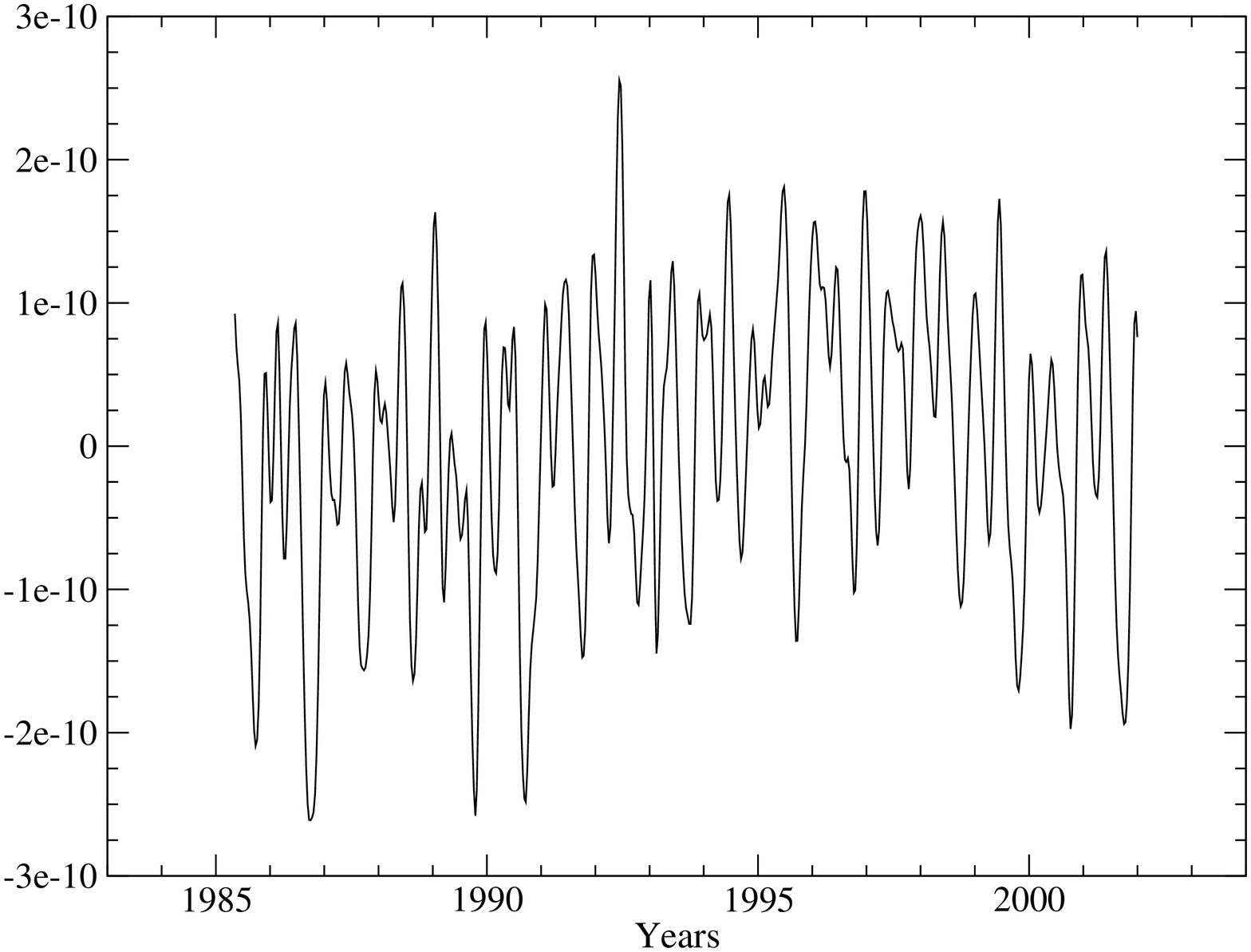}
\caption{Normalized total $\Delta C_{20}$: top is the total series including atmospheric, oceanic tides and solid earth tides effects and the residuals; bottom is the total series without the solid earth tides effect.}
\label{fig:C20_total}
\end{figure}

$\bullet$ Finally, we must consider a series including all the effects described before. Hence, we add them back to the residuals (Fig.~\ref{fig:C20_grgs}), interpolating and filtering the data. Then we obtain the total series (Fig.~\ref{fig:C20_total}). \\
 
\subsection{Adjustments in $\Delta H$ data}

Eq.~(\ref{eq:Delta_Hd}) allows us to transform the geodetic $\Delta \bar{C}_{20}$ temporal variations into the dynamical flattening variations $\Delta H$. They can then be introduced into the precession equations (\ref{eq:Syst_Prec_Nut}), replacing $H$ with $(H + \Delta H)$ (Eq.~(\ref{eq:luni_solar_effect}), Eq.~(\ref{eq:Km}), Eq.~(\ref{eq:Ks}) and Eq.~(\ref{eq:r0})) and using the process already described in Sect. 2.3.

It is generally considered that VLBI observations of the Earth's orientation in space are not sensitive to the atmospheric and oceanic contributions to the variations in $C_{20}$ (de Viron, 2004). However the amplitudes of these effects have been evaluated in Table \ref{tab:Periodic_results} for further discussion and in any case we can notice that they have a negligible effect on precession.

The analytical and semi-analytical approach to solving the precession-nutation equations provides polynomial developments of the $\psi_A$ and $\omega_A$ quantities. The $\Delta H$ data are then considered as a linear expression plus Fourier terms with periods derived from a spectral analysis (18.6-yr, 9.3-yr, annual and semi-annual terms) (see Tables \ref{tab:adjustments_constant_part}, \ref{tab:adjustments}, \ref{tab:adjustments_residus} and \ref{tab:adjustments_total}). Note that the phase angles used for adjusting the $\Delta H$ periodic terms are those of the corresponding nutation terms. This implies changes in the development of the equatorial precession angles ($\psi_A$, $\omega_A$), which we describe in the next section.

For the residual contribution of $\Delta H$, we will consider (i) an adjustment of a secular trend over the interval from 1985 to 1998 (see Table \ref{tab:adjustments_residus} and Eq.~(\ref{eq:Delta_H_res})), and (ii) an adjustment of a secular trend plus a $18.6$-yr periodic term (see Table \ref{tab:adjustments} and Eq.~(\ref{eq:Delta_H_res_18.6})), both added to the seasonal terms. The fit (i) of the secular trend gives:  
\begin{equation}\label{eq:Delta_H_res}
\dot{H} \simeq -7.4 \times 10^{-9} \mbox{/cy} ~\Leftrightarrow ~\dot{J}_2 \simeq -2.5 \times 10^{-9} \mbox{/cy}
\end{equation}
and the fit of model (ii) gives:
\begin{eqnarray}\label{eq:Delta_H_res_18.6}
\Delta H & = & \left( 74 \times 10^{-11} \right) \times t + \left( 20.9 \times 10^{-11} \right) \times \sin(\omega t)    \nonumber \\
         &   &  + \left( 32.5 \times 10^{-11} \right) \times \cos(\omega t) 
\end{eqnarray}
with $\omega = 2 \pi/0.186$.

We must recall that these adjustments have been made together with the fit of annual and semi-annual terms. In contrast, the higher frequency terms appearing in the $\Delta H$ data have been filtered and we therefore did not take into account other contributions, as for example the diurnal effects of the geophysical contributions in $\Delta H$.


\begin{table}
\centering
\caption{Summary of the constant parts for $H$ and $C_{20}$ (Constant part + Permanent tide) used in this study.}
\label{tab:adjustments_constant_part}
\begin{tabular}{lr}
\hline
\hline
\noalign{\smallskip}
 $H_{\mbox{\scriptsize MHB}}$  & $3.2737942 \times 10^{-3}$	 \\ 
\noalign{\smallskip}
\hline
\noalign{\smallskip}
$\bar{C}_{20}$          & $-4.841695 \times 10^{-4}$	 \\ 
\noalign{\smallskip}
$J_2$                   & $1.0826358 \times 10^{-3}$	 \\ 
\noalign{\smallskip}
 $H^*$                  & $3.2671524 \times 10^{-3}$	 \\ 
\noalign{\smallskip}
$H^*$ with geophysical  & $3.2671521 \times 10^{-3}$     \\ 
constant parts          &                                \\
\noalign{\smallskip}
 $H^{**}$               & $3.2735556 \times 10^{-3}$	 \\
\noalign{\smallskip}
\hline
\end{tabular}
\end{table}

\begin{table*}
\centering
\caption{Adjustment of periodic terms in the $\Delta H$ contributions, for the data span 1985-2002 for various $\Delta H$ geophysical sources (atmospheric $\Delta H_{\mbox{\scriptsize ~atm.}}$, oceanic tides $\Delta H_{\mbox{\scriptsize ~oc.}}$ and solid earth tides $\Delta H_{\mbox{\scriptsize ~soltid.}}$, as well as the residuals $\Delta H_{\mbox{\scriptsize ~res.}})$) - Units are in $10^{-10}$ rad.}
\label{tab:adjustments}
\begin{tabular}{lrrcrrcrrcrr}
\hline
\hline
\noalign{\smallskip}
 period   & \multicolumn{2}{c}{$\Delta H_{\mbox{\scriptsize ~res.}}$} & & \multicolumn{2}{c}{$\Delta H_{\mbox{\scriptsize ~atm.}}$} & & \multicolumn{2}{c}{$\Delta H_{\mbox{\scriptsize ~oc.}}$}  & & \multicolumn{2}{c}{$\Delta H_{\mbox{\scriptsize ~soltid.}}$} \\
(in years)&   sin    &   cos    & &    sin   &  cos     & &    sin    &    cos   & &    sin    &   cos       \\
\noalign{\smallskip}
\hline
\noalign{\smallskip}
1         &  $2.17$  &  $-4.02$ & &  $0.96$  & $-1.66$  & &  $-3.92$  &  $1.41$  & &  $-4.64$  &   $0.89$    \\    
\noalign{\smallskip}
0.5       & $-0.43$  &   $3.71$ & & 	     & 	        & &   $0.76$  &  $1.56$  & &  $-0.34$  &  $28.04$    \\    
\noalign{\smallskip}
18.6      &  $2.09$  &   $3.25$ & &          & 	        & & 	      &	         & &  $-0.46$  & $-26.29$    \\
\noalign{\smallskip}
9.3       & $-0.17$  &  $-1.65$ & & 	     & 	        & & 	      &	         & &  $-0.01$  &   $0.29$    \\
\noalign{\smallskip}
\hline
\end{tabular}
\end{table*}

\begin{table}
\centering
\caption{Specific adjustment of the $\Delta H$ residual series ($\Delta H_{\mbox{\scriptsize ~res.}}$), from 1985 to 1998. The secular trend is considered as in Eq.~(\ref{eq:Delta_H_res}) - Units are in $10^{-10}$ rad.}
\label{tab:adjustments_residus}
\begin{tabular}{lrr}
\hline
\hline
\noalign{\smallskip}
 period       & \multicolumn{2}{c}{$\Delta H_{\mbox{\scriptsize ~res.}}$}  \\
(in years)    &    sin    &   cos       \\
\noalign{\smallskip}
\hline
\noalign{\smallskip}
1             &  $2.57$   &  $-3.84$    \\    
\noalign{\smallskip}
0.5           & $-0.50$   &  $ 3.80$    \\    
\noalign{\smallskip}
\hline
\end{tabular}
\end{table}

\begin{table}
\centering
\caption{Adjustment of the total series of $\Delta H$ ($\Delta H_{\mbox{\scriptsize ~tot.}}$), from 1985 to 2002 - Units are in $10^{-10}$ rad.}\label{tab:adjustments_total}
\begin{tabular}{lrr}
\hline
\hline
\noalign{\smallskip}
period    &\multicolumn{2}{c}{$\Delta H_{\mbox{\scriptsize ~tot.}}$} \\ 
(in years)&  sin     &   cos	  \\   
\noalign{\smallskip}
\hline
\noalign{\smallskip}
1         &  $-5.39$ &  $-3.39$   \\ 
\noalign{\smallskip}
0.5       &   $0.07$ &  $33.67$   \\
\noalign{\smallskip}
18.6      &   $0.92$ & $-23.11$   \\  
\noalign{\smallskip}
9.3       &  $-0.08$ &  $-0.50$   \\  
\noalign{\smallskip}
\hline
\end{tabular}
\end{table}


\section{Effects of the $\Delta H$ contributions on the precession angles}

On the basis of the models fitted to the time series of $\Delta H$ in the previous section, obtained with geodetic $\Delta {C}_{20}$ series, we investigate the influence of these geodetic data on the precession angle developments. First, we evaluate the effect of the secular trend considered in the $\Delta \bar{C}_{20}$ residual series. Second, we report on the influence of each geophysical contribution, on the influence of the residuals and on that of the total contribution. Finally we focus on the periodic effects resulting from the various $\Delta H$ contributions.

\subsection{$\dot{J}_2$ influence}

We have already mentioned that the $\dot{J}_2$ influence was taken into account in previous precession solutions (Williams 1994, Capitaine et al. 2003) (see Sect. 4.1). But depending on the value adopted, the polynomial development of the $\psi_A$ precession angle is different. Indeed, if we take $\dot{J}_2 = -2.5 \times 10^{-11}$/cy like in our study, or $\dot{J}_2 = -3 \times 10^{-11}$/cy like in Capitaine et al. (2003), the contribution in $\psi_A$ varies by about $1.5$ mas/cy$^2$ (see Table \ref{tab:J2_rate}). So we must carefully take into account this $J_2$ rate. Furthermore, (i) we already noticed that such a secular trend has been recently discussed because of the change in this trend in 1998 (see Fig.~\ref{fig:C20_grgs}) and (ii) the uncertainty in this secular trend, derived from space measurements of $J_2$, is significant. Therefore we can conclude that until there is a better determination of the $J_2$ rate, the accuracy of the precession expression is limited to about $1.5$ mas/cy$^2$.

\begin{table*}
\centering
\caption{Influence of $\dot{J}_2$ on the polynomial development of $\psi_A$ (more particularly on the $t^2$ and $t^3$ terms): (1) IAU2000 (Mathews et al., 2002), (2) P03 (Capitaine et al., 2003) and (3) Same computation as in P03 but with other $\dot{J}_2$ values. The $J_2$ secular trend estimation based on our $C_{20}$ residuals series is: $\dot{J}_2 = -2.5 \times 10^{-9}$/cy.}\label{tab:J2_rate}
\begin{tabular}{lcrr}
\hline
\hline
\noalign{\smallskip}
                &  $\dot{J}_2$              &    $t^2$     &  $t^3$       \\
\noalign{\smallskip}
\hline 
\noalign{\smallskip}
(1) IAU2000     & None                      & $-1''.07259$  & $-0''.001147$  \\
\noalign{\smallskip}
\hline 
\noalign{\smallskip}
(2) P03         &  $-3 \times 10^{-9}$/cy   & $-1''.079007$ & $-0''.001140$  \\
\noalign{\smallskip}
\hline 
\noalign{\smallskip}
\noalign{\smallskip}
                &      & \multicolumn{2}{c}{$\overbrace{~~~~~~~~~~~~~~~~~~~~~~~~~~~~~~~~~~}^{\mbox{Differences wrt P03}}$}\\
                &  $~~~0$ /cy               &  -7.000 mas  &   2 $\mu$as  \\
                &  $-2 \times 10^{-9}$/cy   &  -2.871 mas  &   1 $\mu$as  \\
(3)             &  $-2.3 \times 10^{-9}$/cy &  -1.954 mas  &   1 $\mu$as  \\
                &  $-2.5 \times 10^{-9}$/cy &  -1.495 mas  &   1 $\mu$as  \\
\noalign{\smallskip}
\hline
\end{tabular}
\end{table*}


\subsection{Precession}

First, we can compare the polynomial part of our solution Geod04 for the precession angles, based on the constant part $H_{MHB}$ of $H$ and on its variable part provided by expression (\ref{eq:Delta_H_res_18.6}), with previous precession expressions (IAU2000 and P03) (see Table \ref{tab:Fourier_results}). The differences larger than one $\mu$as concern the $\psi_A$ precession angle and more particularly its $t^2$ and $t^3$ terms. The differences (of 7 mas and 2 $\mu$as, respectively) with respect to P03 are due to considering or not considering the $\dot{J}_2$ effect. Actually, P03 includes a $J_2$ secular trend, whereas Geod04 includes instead a $18.6$-yr periodic term (see (3) in Table \ref{tab:Fourier_results} or (2) in Table \ref{tab:Geod04}). Comparing Geod04 with the IAU~2000 precession (which does not consider the $J_2$ rate) shows differences of 0.6~mas and 5~$\mu$as in the $t^2$ and $t^3$ terms, respectively. This results from the improved dynamical consistency of the Geod04 solution (based on the P03 precession equations) with respect to IAU~2000. Note that such results regarding the $t^2$ and $t^3$ terms will not be affected if changes of the order of 1 mas/cy in the precession rate would occur in an updated P03 solution.  

Second, we can evaluate the differences introduced in the $\psi_A$ (and $\omega_A$) polynomial development by the use of a constant part for $H$ determined with the geodetic $J_2$ (as used in {\it Geod04-H*} and {\it Geod04-H**}) instead of the $H_{astro}$ determined by VLBI and used in {\it Geod04}. Table \ref{tab:Fourier_results} shows that the differences are very large, but it should be noted that using $J_2$ for deriving $H$ suffers from the too large errors introduced by the mismodelled $C/M{R_e}^2$.


\subsection{Periodic contributions}

On the basis of the adjustments made in Sect.~4.3 for the different $\Delta H$ contributions, we estimate here the periodic effects appearing in the expressions of the precession angles. We can focus on the Fourier terms in the $\psi_A$ precession angle, which are the most sensitive to the $\Delta H$ effects. The corresponding results are presented in Table \ref{tab:Periodic_results}. 

$\bullet$ First, we note that the major effect is due to the $18.6$-yr periodic term in the solid Earth tides (contribution number~(3) of Table \ref{tab:Periodic_results}): about $-2~\mu$as and $120~\mu$as in cosine and sine, respectively. The tidal annual and semi-annual effects are negligible as well as the atmospheric and oceanic effects (contributions number~(4) and (5) of Table \ref{tab:Periodic_results}).

$\bullet$ Second, note that the $\Delta H$ variation strictly limited to its residual part, not modelled into the geodetic orbit restitution, introduces negligible Fourier terms into the $\psi_A$ development. But we can note that the way the long term effect is considered in such data (i.e.~ either with a secular trend term (contribution number~(1) of Table \ref{tab:Periodic_results}) or a $18.6$-yr periodic term (contribution number~(2) of Table \ref{tab:Periodic_results})) is important. Modelling the long term variation in the geodetic residuals over the total data span as a $18.6$-yr variation induces a term with an amplitude of $15~\mu$as in the $\psi_A$ development. But at present the $\Delta C_{20}$ data span is not long enough to allow us to discriminate between the two models.     

Finally, we can conclude that the geodetic determination of the total variable $C_{20}$ (contribution number~(6) of Table \ref{tab:Periodic_results}) introduces Fourier terms into the $\psi_A$ precession angle development, mainly a $18.6$-yr periodic one, of the order of $4~\mu$as and $105~\mu$as in cosine and sine, respectively.


\begin{table*}
\centering
\caption{Polynomial part of the $\psi_A$ and $\omega_A$ developments (units in arcseconds): comparison of (1)  IAU2000 (Mathews et al. 2002) - (2) P03 (Capitaine et al. 2003) - (3) Differences of Geod04 (this study) with respect to P03, considering all the contributions for $\Delta H$ (Table \ref{tab:adjustments_total}, $\Delta H_{~tot}$) - (4) Differences of Geod04 with respect to P03, obtained with a $H$ constant part different from $H_{MHB}$, but not used in the following (see Table \ref{tab:Coeff_et_H} for the $H^*$ and $H^{**}$ constant values).}\label{tab:Fourier_results}
\centering
\begin{tabular}{llcccc}
\hline
\hline
\noalign{\smallskip}
Angle      & Source            & $t^0$  &       $t$	        &     $t^2$           &  $t^3$            \\
\noalign{\smallskip}
\hline
\noalign{\smallskip}
	   &(1) {\it IAU2000}  &        & {\it $5038''.478750$} & {\it $-1''.07259$}  & {\it $-0''.001147$}  \\
\noalign{\smallskip}
           &(2) {\it P03}      &        & {\it $5038''.481507$} & {\it $-1''.079007$} & {\it $-0''.001140$}  \\
\noalign{\smallskip}
\cline{2-6}
\noalign{\smallskip}
\noalign{\smallskip}
$\psi_A$   &                   &        & \multicolumn{3}{c}{$\overbrace{~~~~~~~~~~~~~~~~~~~~~~~~~~~~~~~~~~~~~~~~~~~~~~~~~~~~~~~~~~~~~~~}^{\mbox{Differences wrt P03}}$}\\
           &(3) Geod04 ~~$H_{MHB}$&	&         $0''$         &      -7 mas         &   2 $\mu$as       \\
\noalign{\smallskip}
\noalign{\smallskip}
	   &(4) Geod04 
	     $\left\{
	     \begin{tabular}{l}
	     $H^*$   \\
	     $H^{**}$               
	     \end{tabular}
	     \right.$          &\begin{tabular}{l}
				  \\
                                   	 
				\end{tabular} &\begin{tabular}{c}
					       $\simeq$ $10''.23$  \\ 
					       $\simeq$ $0''.37$
					       \end{tabular} &\begin{tabular}{l}
							      -9.177 mas  \\
							      -7.079 mas
							      \end{tabular}     &\begin{tabular}{c}
									         -3 $\mu$as \\
									          2 $\mu$as
									         \end{tabular}      \\
\noalign{\smallskip}
\hline
\noalign{\smallskip}
\noalign{\smallskip}
\noalign{\smallskip}
\noalign{\smallskip}
\hline
\hline
\noalign{\smallskip}
	   &(1) {\it IAU2000}& {\it $84381''.448$}  & {\it $-0''.025240$}  & {\it $0''.05127$}   & {\it $-0''.00772$}  \\
\noalign{\smallskip}
           &(2) {\it P03}    & {\it $84381''.406$}  & {\it $-0''.025754$}  & {\it $0''.051262$}  & {\it $-0''.007725$} \\
\noalign{\smallskip}
\cline{2-6}
\noalign{\smallskip}
\noalign{\smallskip}
$\omega_A$ &                 & \multicolumn{4}{c}{$\overbrace{~~~~~~~~~~~~~~~~~~~~~~~~~~~~~~~~~~~~~~~~~~~~~~~~~~~~~~~~~~~~~~~~~~~~~~~~~~~~}^{\mbox{Differences wrt P03 in $\mu$as}}$}\\      
           &(3) Geod04 ~~$H_{MHB}$&        0     &        0          &       0          &     0            \\
\noalign{\smallskip}
\noalign{\smallskip}
	   &(4) Geod04 
	     $\left\{
	     \begin{tabular}{l}
	     $H^*$   \\
	     $H^{**}$               
	     \end{tabular}
	     \right.$       &\begin{tabular}{l}
			      0 \\
                              0	
			     \end{tabular}      &\begin{tabular}{l}
					          0 \\ 
					          0
					         \end{tabular}     &\begin{tabular}{c}
							             104   \\
							               3   
							            \end{tabular}     &\begin{tabular}{l}
									               -35   \\
									               ~-1 
									               \end{tabular}      \\
\noalign{\smallskip}
\hline
\end{tabular}
\end{table*}


\begin{table*}
\centering
\caption{Polynomial part of the $\psi_A$ development (units in arcseconds) for various $\Delta H$ sources used in our study, with respect to P03: comparison of (1) P03 (Capitaine et al., 2003) - (2) Difference between P03 and Geod04 (i.e. the effect of the total $\Delta H$) - (3) Difference between P03 and the effect of the $\Delta H$ residuals.}\label{tab:Geod04}
\centering
\begin{tabular}{llccc}
\hline
\hline
\noalign{\smallskip}
Angle      &   Source                   &     $t$               &     $t^2$           &     $t^3$         \\
\noalign{\smallskip}
\hline
\noalign{\smallskip}
           &(1) {\it P03}               & {\it $5038''.481507$} & {\it $-1''.079007$} & {\it $-0''.001140$}  \\
\noalign{\smallskip}
\cline{2-5}
\noalign{\smallskip}
\noalign{\smallskip}
$\psi_A$   &                            & \multicolumn{3}{c}{$\overbrace{~~~~~~~~~~~~~~~~~~~~~~~~~~~~~~~~~~~~~~~~~~~~~~~~~~~~~~~~~~}^{\mbox{Differences wrt P03 in $\mu$as}}$}\\ 
           &(2) Geod04 total contributions &        0           &      -7000          &      2            \\
\noalign{\smallskip}
	   &(3) Geod04 residuals 
	     $\left\{
	     \begin{tabular}{l}
	     1985-1998   \\
	     1985-2002              
	     \end{tabular}
	     \right.$                   &\begin{tabular}{l}
					 0 \\ 
					 0
					 \end{tabular}&\begin{tabular}{c}
						       -1495 \\
						       -7000 
						       \end{tabular}&\begin{tabular}{l}
							             1   \\
								     2  
								     \end{tabular} \\
\noalign{\smallskip}
\hline
\end{tabular}
\end{table*}


\begin{table*}
\caption{Fourier part of the $\psi_A$ development, depending on the contribution considered for the $\Delta H$ periodic effect (units in $\mu$as).}\label{tab:Periodic_results}
\centering
\begin{tabular}{lcllrr}
\hline
\hline
\noalign{\smallskip}
\multicolumn{6}{r}{Periodic contribution for the $t^0$ term of $\psi_A$} \\
\noalign{\smallskip}
\hline
\noalign{\smallskip}
\multicolumn{4}{c}{}                             &  \multicolumn{2}{c}{$\mu$as}  \\
\noalign{\smallskip}
\multicolumn{4}{c}{}                             &      cos       &      sin     \\
\noalign{\smallskip}
\hline
\noalign{\smallskip}
              & (1) &  Residuals   & Annual      &     -1 	  &    -1         \\
              &     & (1985-1998)  & Semi-annual &      - 	  &     1         \\  
\noalign{\smallskip}
\noalign{\smallskip}
              & (2) &  Residuals   & 18.6-yr     &     10         &   -15        \\   
              &     & (1985-2002)  & 9.3-yr      &      -         &     4        \\ 
              &     &              & Annual      &     -1         &    -1        \\ 
              &     &              & Semi-annual &      -         &     1        \\ 
\noalign{\smallskip}
\noalign{\smallskip}
$\Delta H$    & (3) & Solid tides  & 18.6-yr	 &	-2	  &   120	 \\   
periodic      &     &              & 9.3-yr	 &	 -	  &    -1	 \\  
contributions &     &              & Annual      &       1        &     -        \\ 
              &     &              & Semi-annual &       -        &     3        \\  
\noalign{\smallskip}
\cline{2-6}
\noalign{\smallskip}
              & (4) & Ocean tides  & Annual      &      1         &     -        \\  
              &     &              & Semi-annual &      -         &     -        \\ 
\noalign{\smallskip}
\noalign{\smallskip}
              & (5) & Atmosphere   & Annual      &      -         &     -        \\ 
\noalign{\smallskip}
\cline{2-6}
\cline{2-6}
\noalign{\smallskip}
              & (6) & Geod04        & 18.6-yr     &     4         &   105        \\  
              &     & total         & 9.3-yr      &     -         &     1        \\  
              &     & contributions & Annual      &     1         &    -1        \\
              &	    &               & Semi-annual &     -         &     4        \\	
\noalign{\smallskip}
\hline
\end{tabular}
\end{table*}


\section{Discussion} 

This study was based on new considerations: the use of a geodetic determination of the variable geopotential to investigate its influence on the developments of the precession angles. 
The major effect on the precession is due to the $J_2$ secular trend which implies an acceleration of the $\psi_A$ precession angle. But for the moment, the available time span for $J_2$ satellite series is not as long as we need to determine a reliable $\dot{J}_2$ value. The $J_2$ secular trend estimation based on our $C_{20}$ residuals series from 1985 to 1998 is: $\dot{J}_2 = -2.5 \times 10^{-9}$/cy. The accuracy of the precession expression is limited to about $1.5$ mas/cy$^2$ due to the uncertainty in this $J_2$ rate value.

Then, we can notice that the main periodic effect is due to the $18.6$-yr periodic term in $\Delta C_20$ due to solid Earth tides. But we must say that computing the $\Delta C_{20}$ with satellite positioning observations requires making some assumptions on the geophysical contributions to $\Delta C_{20}$, for instance from atmospheric pressure, and oceanic or solid Earth tides. Actually, models are used, but they are not perfect and we may have some errors. So the $\Delta C_{20}$ residuals obtained may be affected by these errors, which is why the total $\Delta C_{20}$ contributions (residuals observed + models assumed) constitute a better series to evaluate the effects on the precession angles. This introduces Fourier terms into the $\psi_A$ development ($4~\mu$as and $105~\mu$as in cosine and sine respectively; see Table \ref{tab:Periodic_results}) that we should compare to the MHB2000-nutations.
Indeed, the different terms of the total $\Delta H$ (or $\Delta C_{20}$) contributions have same periods as the ($\Delta \psi$, $\Delta \epsilon$) nutations. This implies that there is some coupling between the observed $\Delta H$ effects and the nutations, which may not have been included in the MHB2000-nutations.   

In the future, we will be able to compare the $J_2$ data with geophysical models and data, in order to have better ideas on the different contributions and on the secular trend.
We will also be able to proceed to numerical study of this problem, and to implement a refined and more realistic Earth model.


\begin{acknowledgements}
We are grateful to V. Dehant for helpful advice and information. We thank C. Bizouard, O. de~Viron, S. Lambert, and J. Souchay, for valuable discussion and J. Chapront for providing the well documented software GREGOIRE. We also thank the referee for valuable suggestions for improving the presentation of the manuscript.
\end{acknowledgements}


\end{document}